\documentclass[onecolumn,10pt,oneside,draftclsnofoot]{IEEEtran}%

\usepackage{amsbsy}
\usepackage{floatflt} 

\usepackage{amsmath}
\usepackage{amssymb}
\usepackage{times}
\usepackage{graphicx}
\usepackage{xspace}
\usepackage{paralist} 
\usepackage{setspace} 
\usepackage{xypic}
\xyoption{curve}
\usepackage{latexsym}
\usepackage{theorem}
\usepackage{ifthen}
\usepackage{subfigure}
\usepackage[ruled,vlined]{algorithm2e}
\usepackage{booktabs}

{\theoremheaderfont{\it} \theorembodyfont{\rmfamily}
\newtheorem{theorem}{Theorem}
\newtheorem{lemma}[theorem]{Lemma}
\newtheorem{definition}[theorem]{Definition}

\newtheorem{corollary}[theorem]{Corollary}

\newtheorem{proposition}[theorem]{Proposition}

}

\renewcommand{\baselinestretch}{1.4}

\title{Distributed Consensus Algorithms in Sensor Networks: Quantized
Data and Random Link Failures}
\author{Soummya Kar and Jos\'e M.~F.~Moura$^{*}$
\thanks{The authors are with the Department of Electrical and Computer Engineering,
Carnegie Mellon University, Pittsburgh, PA, USA 15213 (e-mail:
soummyak@andrew.cmu.edu, moura@ece.cmu.edu, ph: (412)268-6341,
fax: (412)268-3890.)}
\thanks{Work supported by NSF under grants \#~ECS-0225449
and~\#~CNS-0428404, by an IBM Faculty Award, and by  the Office of Naval Research under MURI N000140710747.}
\thanks{Manuscript initially submitted on December 10, 2007; last revised on July 19, 2009.}}

\begin{document}
\maketitle \thispagestyle{empty} \maketitle

\begin{abstract}The paper studies the problem of distributed average consensus in sensor networks with quantized data and random link
failures.  To achieve consensus, dither (small noise) is added to the sensor states before quantization. When the quantizer range is unbounded (countable number of quantizer levels), stochastic approximation shows that consensus is asymptotically achieved with probability one and in mean square to a finite random variable. We show that the mean-squared error (m.s.e.) can be made arbitrarily small by tuning the link weight sequence, at a cost of the convergence rate of the algorithm. To study dithered consensus with random links when the range of the quantizer is bounded, we establish uniform boundedness of the sample paths of the unbounded quantizer. This requires characterization of the statistical properties of the supremum taken over the sample paths of the state of the quantizer. This is accomplished by splitting the state vector of the quantizer in two components: one along the consensus subspace and the other along the subspace orthogonal to the consensus subspace.  The proofs use maximal inequalities for submartingale and supermartingale sequences. From these, we derive probability bounds on the excursions of the two subsequences, from which probability bounds on the excursions of the quantizer state vector follow. The paper shows how to use these probability bounds to design the quantizer parameters and to explore tradeoffs among the number of quantizer levels, the size of the quantization steps, the desired probability of saturation, and the desired level of accuracy~$\epsilon$ away from consensus. Finally, the paper illustrates the quantizer design with a numerical study.
\end{abstract}

\hspace{.43cm}\textbf{Keywords:} Consensus, quantized, random link
failures, stochastic approximation, convergence, bounded quantizer, sample path behavior, quantizer saturation

\newpage
\section{Introduction}
\label{introduction} This paper is concerned with consensus in
networks, e.g.,  a sensor network, when the data exchanges among
nodes in the network (sensors, agents) are quantized. Before detailing our work, we briefly overview the literature.

\textbf{Literature review.} Consensus is
broadly understood as individuals in a community achieving a
consistent view of the World by interchanging information
regarding their current state with their neighbors. Considered in
the early work of Tsitsiklis \emph{et. al.}
(\cite{tsitsiklisphd84,tsitsiklisbertsekasathans86}), it has
received considerable attention in recent years and arises in
numerous applications including: load balancing, \cite{cybenko89},
alignment, flocking, and multi-agent collaboration, e.g.,
\cite{jadbabaielinmorse03,SensNets:Olfati04},
vehicle formation, \cite{faxmurray04},
 gossip algorithms, \cite{Boyd-GossipInfTheory}, 
  tracking, 
   data fusion, \cite{salig:06}, 
    and distributed inference, \cite{tsp06-K-A-M}. We refer the reader to the recent overviews on consensus, which include \cite{olfatisaberfaxmurray07,renbeardatkins07}.

Consensus is a distributed iterative algorithm where the sensor
states evolve on the basis of local interactions.
Reference~\cite{SensNets:Olfati04} used spectral graph concepts
like graph Laplacian and algebraic connectivity to prove
convergence for consensus under several network operating
conditions (e.g., delays and switching networks, i.e., time
varying). Our own prior work has been concerned with designing
topologies that optimize consensus with respect to the convergence
rate, \cite{Allerton06-K-M,tsp06-K-A-M}. Topology design is
concerned with two issues: \begin{inparaenum}[1)] \item the
definition of the graph that specifies the neighbors of each
sensor---i.e., with whom should each sensor exchange data; and
\item the weights used by the sensors when combining the
information received from their neighbors to update their state.
\end{inparaenum}
Reference~\cite{Boyd} considers the problem of weight design, when the topology is specified, in the framework of semi-definite
programming. References~\cite{Olfati-ConsSmallWorld}, \cite{saeed:05} considered the impact of different topologies on
the convergence rate of consensus, in particular, regular, random, and small-world graphs, \cite{SensNets:Watts98}. Reference~\cite{Barooah-Hespanha} relates the convergence properties of consensus algorithms to the effective resistance of the network, thus obtaining convergence rate scaling laws for networks in up to 3-dimensional space. Convergence results for general problems in multi-vehicle formation has been considered in~\cite{Bamieh}, where convergence rate is related to the topological dimension of the network and stabilizability issues in higher dimensions are addressed. Robustness issues in consensus algorithms in the presence of analog communication noise and random data packet dropouts have been considered in~\cite{karmoura-randomtopologynoise}.

\textit{Review of literature on quantized consensus.}
 Distributed consensus with quantized transmission has been studied recently in~\cite{Kashyap,yildizscaglione07,Tuncer,Fagnani} with respect to time-invariant (fixed) topologies. Reference~\cite{Nedic} considers quantized consensus for a certain class of time-varying topologies. The algorithm in~\cite{Kashyap} is restricted to integer-valued initial sensor states, where at each iteration the sensors exchange integer-valued data. It is shown there that the sensor states are asymptotically close (in their appropriate sense) to the desired average, but may not reach absolute consensus. In~\cite{yildizscaglione07}, the noise in the consensus algorithm studied in~\cite{SensNets:Xiao05} is interpreted as quantization noise and shown there by simulation with a small network that the variance of the quantization noise is reduced as the algorithm iterates and the sensors converge to a consensus. References~\cite{Tuncer,Tuncer-Allerton} study probabilistic quantized consensus. Each sensor updates its state at each iteration by probabilistically quantizing its current state (which~\cite{Tuncer-dither} claims equivalent to dithering) and linearly combining it with the \emph{quantized} versions of the states of the neighbors. They show that the sensor states reach consensus a.s.~to a quantized level. In~\cite{Fagnani} a worst case analysis is presented on the error propagation of consensus algorithms with quantized communication for various classes of time-invariant network topologies, while~\cite{Carli-Fagnani} addresses the impact of more involved encoding/decoding strategies, beyond the uniform quantizer. The effect of communication noise in the consensus process may lead to several interesting phase transition phenomena in global network behavior, see, for example, \cite{Vicsek} in the context of a network of mobile agents with a non-linear interaction model and~\cite{Elia}, which rigorously establishes a phase transition behavior in a network of bipolar agents when the communication noise exceeds a given threshold. Consensus algorithms with general imperfect communication (including quantization) in a certain class of time-varying topologies has been addressed in~\cite{Nedic}, which assumes that there exists a window of fixed length, such that the union of the network graphs formed within that window is strongly connected. From a distributed detection viewpoint, binary consensus algorithms over networks of additive white Gaussian noise channels were addressed in~\cite{Mostofi}, which proposed soft information processing techniques to improve consensus convergence properties over such noisy channels.  The impact of fading on consensus is studied in~\cite{Mostofi-fading}.

\textbf{Contributions of this paper.}
 We consider consensus with \emph{quantized} data and \emph{random} inter-sensor \emph{link failures}. This is useful in applications where limited bandwidth and power  for inter-sensor communications preclude exchanges of high precision (analog) data as in wireless sensor networks. Further, randomness in the environment results in random data packet dropouts.  To handle quantization, we modify standard consensus by adding a small amount of noise, \emph{dither}, to the data before quantization and by letting the consensus weights to be time varying, satisfying a \emph{persistence} condition--their sum over time diverges, while their square sum is finite. We will show that dithered quantized consensus in networks with random links  converges.

The randomness of the network topology is captured by assuming that the time-varying Laplacian sequence, $\{L(i)\}_{i\geq 0}$, which characterizes the communication graph, is independent with mean $\overline{L}$; further, to prove convergence, we will need the mean graph algebraic connectivity (first nonzero eigenvalue of~$\overline{L}$) $\lambda_{2}(\overline{L})>0$, i.e., the network to be connected on the average. Our proofs do not require any distributional assumptions on the link failure model (in space). During the same iteration, the link failures can be spatially dependent, i.e., correlated across different edges of the network. The model we work with in this paper subsumes the erasure network model, where link failures are independent both over space and time. Wireless sensor networks motivate us since interference among the sensors communication correlates the link failures over space, while over time, it is still reasonable to assume that the channels are memoryless or independent. Note that the assumption $\lambda_{2}\left(\overline{L}\right)>0$ does not require the individual random instantiations of $L(i)$ to be connected; in fact, it is possible to have all the instantiations to be disconnected. This captures a broad class of asynchronous communication models, for example, the random asynchronous gossip protocol in~\cite{Boyd-Gossip} satisfies $\lambda_{2}\left(\overline{L}\right)>0$ and hence falls under this framework.

The main contribution of this paper is the study of the convergence and the detailed analysis of the sample path of this dithered distributed quantized consensus algorithm with random link failures. This distinguishes our work from~\cite{Kashyap} that considers fixed topologies (no random links) and integer valued initial sensor states, while our initial states are arbitrarily real valued. To our knowledge, the convergence and sample path analysis of dithered quantized consensus with random links has not been carried out before. The sample path analysis of quantized consensus algorithms is needed because in practice quantizers work with bounded (finite) ranges. The literature usually pays thrift attention or simply ignores the boundary effects induced by the bounded range of the quantizers; in other words, although assuming finite range quantizers, the analysis in the literature ignores the boundary effects. Our paper studies carefully the sample path behavior of quantized consensus when the range of the quantizer is bounded. It computes, under appropriate conditions, the probability of large excursion of the sample paths and shows that the quantizer can be designed so that with probability as close to~$1$ as desired the sample path excursions remain bounded, within an $\epsilon$-distance of the desired consensus average. Neither our previous work~\cite{karmoura-randomtopologynoise}, which deals with consensus with noisy analog communications in a random network, nor references~\cite{Tuncer,Tuncer-Allerton,Tuncer-dither}, which introduce a probabilistic quantized consensus algorithm in fixed networks,  nor~\cite{Mesbahi-Noise}, which studies consensus with analog noisy communication and fixed network, study the sample path behavior of quantized consensus. Also, while the probabilistic consensus in~\cite{Tuncer,Tuncer-Allerton,Tuncer-dither} converges almost surely to a quantized level, in our work, we show that dithered consensus converges a.s.~to a random variable which can be made arbitrarily close to the desired average.

To study the a.s.~convergence and m.s.s.~convergence of the dithered distributed quantizers with random links and \emph{unbounded} range, the stochastic approximation method we use in~\cite{karmoura-randomtopologynoise} is sufficient. In simple terms,  we associate, like in~\cite{karmoura-randomtopologynoise}, with the quantized distributed consensus a Lyapounov function and study the behavior of this Lyapounov function along the trajectories of the noisy consensus algorithm with random links. To show almost sure convergence, we show that a functional of this process is a nonnegative supermartingale; convergence follows from convergence results on nonnegative supermartingales. We do this in Section~\ref{quantizedconsensus} where we term the \emph{unbounded} dithered distributed quantized consensus algorithm with random links simply Quantized Consensus, for short, or QC~algorithm. Although the general principles of the approach are similar to the ones in~\cite{karmoura-randomtopologynoise}, the details are different and not trivial--we minimize the overlap and refer the reader to~\cite{karmoura-randomtopologynoise} for details. A second reason to go over this analysis in the paper for the QC algorithm is that  we derive in this Section for QC several specific bounds that are used and needed as intermediate results for the sample path analysis that is carried out in Section~\ref{QCFmajor} when studying dithered quantized consensus when the quantizer is bounded, i.e., Quantized Consensus with Finite quantizer, the QCF quantizer.
 The QCF is a very simple algorithm: it is QC till the QC state reaches the quantizer bound, otherwise an error is declared and the algorithm terminated. To study QCF, we establish uniform boundedness of the sample paths of the QC algorithm. This requires establishing the statistical properties of the supremum taken over the sample paths of the QC. This is accomplished by splitting the state vector of the quantizer in two components: one along the consensus subspace and the other along the subspace orthogonal to the consensus subspace.  These proofs use maximal inequalities for submartingale and supermartingale sequences. From these, we are able to derive probability bounds on the excursions of the two subsequences, which we use to derive probability bounds on the excursions of the QC. We see that to carry out this sample path study requires new methods of analysis that go well beyond the stochastic approximation methodology that we used in our paper~\cite{karmoura-randomtopologynoise}, and also used by~\cite{Mesbahi-Noise} to study consensus with noise but fixed networks. The detailed sample path analysis leads to bounds on the probability of the sample path excursions of the QC algorithm. We then use these bounds to design the quantizer parameters and to explore tradeoffs among these parameters. In particular, we derive a probability of $\epsilon$-consensus expressed in terms of the (finite) number of quantizer levels, the size of the quantization steps, the desired probability of saturation, and the desired level of accuracy~$\epsilon$ away from consensus.

For the QC algorithm, there exists an interesting trade-off between the m.s.e.~(between the limiting random variable and the desired initial average) and the convergence: by tuning the link weight sequence appropriately, it is possible to make the m.s.e.~arbitrarily small (irrespective of the quantization step-size), though penalizing the convergence rate. To tune the QC-algorithm, we introduce a scalar control parameter~$s$ (associated with the time-varying link weight sequence), which can make the m.s.e. as small as we want, irrespective of how large the step-size $\Delta$ is. This is significant in applications that rely on accuracy and may call for very small m.s.e. for being useful.  More specifically, if a cost structure is imposed on the consensus problem, where the  objective is a function of the m.s.e. and the convergence rate, one may obtain the optimal scaling $s$ by minimizing the cost from the Pareto-optimal curve generated by varying~$s$. These tradeoffs and vanishingly small m.s.e. contrasts with the algorithms in~\cite{Kashyap,yildizscaglione07,Tuncer,Fagnani,Nedic} where the m.s.e.~is proportional to $\Delta^{2}$, the quantization step-size--if the step-size is large, these algorithms lead to a large m.s.e.


\textbf{Organization of the paper.} We comment briefly on the organization of the main sections of the paper. Section~\ref{randomtopology} 
summarizes relevant background, including spectral graph theory
and average consensus, and presents the dithered quantized
consensus problem with the dither satisfying the Schuchman
conditions.
 Sections~\ref{QCmajor} considers the convergence of the~QC algorithm. It shows a.s.~convergence to a random variable, whose m.s.e.~is fully characterized. Section~\ref{QCFmajor} studies the sample path behavior of the QC algorithm through the QCF. It uses the expressions we derive for the probability of large excursions of the sample paths of the quantizer to consider the tradeoffs among different quantizer parameters, e.g., number of bits and quantization step, and the network topology to achieve optimal performance under a constraint on the number of levels of the quantizer. These tradeoffs are illustrated with a numerical study.
Finally, Section~\ref{conclusion} concludes the paper.

\section{Consensus with Quantized Data: Problem Statement}
\label{randomtopology} \label{sec:consensusquantized} We
present preliminaries needed for the analysis of the
consensus algorithm with quantized data. The set-up of the average consensus problem is standard,
see the introductory sections of relevant recent papers.

\subsection{Preliminaries: Notation and Average Consensus}
\label{preliminaries}
The sensor network at time index $i$
is represented by an undirected, simple, connected graph
$G(i)=(V,E(i))$. The vertex and edge sets~$V$ and~$E(i)$, with
cardinalities $|V|=N$ and $|E(i)|=M(i)$, collect the sensors and
communication channels or links among sensors in the network at
time $i$. The network topology at time $i$, i.e., with which
sensors does each sensor communicate with, is described by the
$N\times N$ discrete Laplacian $L(i)=L^T(i)=D(i)-A(i)\ge0$. The
matrix~$A(i)$ is the adjacency matrix of the connectivity graph at
time $i$, a $(0,1)$ matrix where $A_{nk}(i)=1$ signifies that
there is a link between sensors $n$ and $k$ at time $i$. The
diagonal entries of~$A(i)$ are zero. The diagonal matrix~$D(i)$ is
the degree matrix, whose diagonal $D_{nn}(i)=d_n(i)$ where
$d_n(i)$ is the degree of sensor~$n$, i.e., the number of links of
sensor~$n$ at time $i$. The neighbors of a sensor or node~$n$,
collected in the neighborhood set $\Omega_n(i)$, are those
sensors~$k$ for which entries $A_{nk}(i)\neq 0$. The Laplacian is positive semidefinite; in case the network
is connected at time $i$, the corresponding algebraic connectivity
or Fiedler value is positive, i.e., the second eigenvalue of the
Laplacian $\lambda_2(L(i))>0$, where the eigenvalues of~$L(i)$ are
ordered in increasing order. For detailed treatment of graphs and
their spectral theory see, for example,
\cite{FanChung,Mohar,SensNets:Bollobas98}. Throughout the paper the symbols $\mathbb{P}[\cdot]$ and $\mathbb{E}[\cdot]$ denote the probability and expectation operators w.r.t. the probability space of interest.

{\bf Distributed Average Consensus.}
\label{distavgcons} The sensors measure
the data $x_n(0)$, $n=1,\cdots,N$, collected in the
vector $\mathbf{x}(0)=\left[x_{1}(0)\cdots x_{N}(0)\right]^{T} \in
\mathbb{R}^{N\times1}$. Distributed average consensus
computes the average~$r$ of the data
\begin{eqnarray}
\label{def_r} r&=&x_{\mbox{\scriptsize{avg}}}(0)=\frac{1}{N}\sum_{n=1}^{N}x_{n}(0)
\label{def:rb}
        = \frac{1}{N}\mathbf{x}(0)^T\mathbf{1}
\end{eqnarray}
by local data exchanges among neighboring sensors.
In~(\ref{def:rb}), the column vector~$\mathbf{1}$ has all entries
equal to~$1$. Consensus is an iterative algorithm where at
iteration~$i$ each sensor updates its current state $x_n(i)$ by a
weighted average of its current state and the states of its
neighbors. Standard consensus assumes a fixed connected network
topology, i.e., the links stay online permanently, the
communication is noiseless, and the data exchanges are analog.
Under mild conditions, the states of all sensors reach consensus,
converging to the desired average~$r$, see
\cite{SensNets:Olfati04,Boyd},
\begin{equation}
\label{eqn:consensuslimit}
\lim_{i\rightarrow\infty}\mathbf{x}(i)=r\mathbf{1}
\end{equation}
where $\mathbf{x(i)}=\left[x_1(i)\cdots x_N(i)\right]^T$ is the
state vector that stacks the state of the~$N$ sensors at
iteration~$i$. We consider consensus with quantized data exchanges and random topology (links fail or become alive at
random times), which models packet dropouts.
In~\cite{karmoura-randomtopologynoise}, we studied consensus
with random topologies and (analog) noisy communications.

\subsection{Dithered Quantization: Schuchman Conditions}
\label{ditheredquantization} We write the sensor updating
equations for consensus with quantized data and random
link failures as
\begin{equation}
\label{genstateupdate}
x_{n}(i+1)=\left[1-\alpha(i)d_n(i)\right]x_{n}(i)+
\alpha(i)\sum_{l\in\Omega_{n}(i)}f_{nl,i}\left[x_{l}(i)\right],
\:\: 1\leq n\leq N
\end{equation}
where: $\alpha(i)$ is the weight at iteration~$i$; and
$\{f_{nl,i}\}_{1\leq n,l\leq N,~i\geq 0}$ is a sequence of
functions (possibly random) modeling the quantization effects.
Note that in~(\ref{genstateupdate}), the weights $\alpha(i)$ are
the same across all links---the equal weights consensus,
see~\cite{Boyd}---but the weights may change with time. Also, the
degree $d_n(i)$ and the neighborhood $\Omega_n(i)$ of each
sensor~$n$, $n=1,\cdots, N$ are dependent on~$i$ emphasizing the
topology may be random time-varying.


{\bf Quantizer.} Each inter-sensor communication channel uses a
uniform quantizer with quantization step $\Delta$. We model the
communication channel by introducing the quantizing function,
$q(\cdot):\mathbb{R}\rightarrow\mathcal{Q}$,
\begin{equation}
\label{qfuncdef} q(y) = k\Delta, \:\:\:(k-\frac{1}{2})\Delta\leq
y<(k+\frac{1}{2})\Delta
\end{equation}
where $y\in\mathbb{R}$ is the channel input. Writing
\begin{equation}
\label{quanterror} q(y)=y+e(y)
\end{equation}
where $e(y)$ is the quantization error. Conditioned on the input, the quantization error~$e(y)$ is deterministic, and
\begin{equation}
\label{quanterror1} -\frac{\Delta}{2}\leq e(y)<\frac{\Delta}{2},
\:\:\:\forall y
\end{equation}
We first consider quantized consensus~(QC) with unbounded range, i.e., the quantization alphabet
 \begin{equation}
\label{Qdef} \mathcal{Q} = \left\{k\Delta\left|\right.k\in\mathbb{Z}\right\}
\end{equation}
 is countably infinite. In Section~\ref{QCFmajor}. we consider what happens when the range of the quantizer is
 finite--quantized consensus with finite~(QCF) alphabet. This study requires that we detail the sample path behavior of the QC-algorithm.

We discuss briefly why a naive approach to consensus will fail
(see~\cite{Tuncer-dither} for a similar discussion.)
 If we use directly the quantized state information, the functions $f_{nl,i}(\cdot)$ in eqn.~(\ref{genstateupdate}) are
\begin{eqnarray}
\label{quantfunc}
f_{nl,i}(x_{l}(i))&=&q(x_{l}(i))\\
\label{quantfunc-b}
    &=&x_{l}(i)+e(x_{l}(i))
\end{eqnarray}
Equations~(\ref{genstateupdate}) take then the form
\begin{equation}
\label{quantupdate}
x_{n}(i+1)=\left[(1-\alpha(i)d_{n}(i))x_{n}(i)+
\alpha(i)\sum_{l\in\Omega_{n}(i)}x_{l}(i)\right]+
\alpha(i)\sum_{l\in\Omega_{n}(i)}e(x_{l}(i))
\end{equation}
The non-stochastic errors (the most right terms
in~(\ref{quantupdate})) lead to error accumulation. If the network
topology remains fixed (deterministic topology,) the update in
eqn.~(\ref{quantupdate}) represents a sequence of iterations that,
as observed above, conditioned on the initial state, which then determines the input, are
deterministic. If we choose the weights $\alpha(i)$'s to decrease
to zero very quickly, then~(\ref{quantupdate}) may terminate
before reaching the consensus set. On the other hand, if the
$\alpha(i)$'s decay slowly, the quantization errors may
accumulate, thus making the states unbounded.

In either case, the naive approach to consensus with quantized
data fails to lead to a reasonable solution. This failure is due to the fact that the error terms are not
stochastic. To overcome these problems, we introduce in

a controlled way noise (dither) to randomize the sensor states
prior to quantizing the perturbed stochastic state. Under appropriate conditions, the resulting quantization
errors possess nice statistical properties, leading to the
quantized states reaching consensus (in an appropriate sense to be
defined below.) Dither places consensus with quantized data in the
framework of distributed consensus with noisy communication links; when the range of the quantizer is unbounded,
we apply stochastic approximation to study the
limiting behavior of QC, as we did in~\cite{karmoura-randomtopologynoise} to study consensus with (analog) noise and random topology. Note that if instead of adding dither, we assumed that the quantization errors are independent, uniformly distributed random variables, we would not need to add dither, and our analysis would still apply.

{\bf Schuchman conditions.}
\label{statquant}
The dither added to randomize the quantization effects  satisfies a special condition, namely, as in subtractively dithered systems, see~\cite{Lipshitz,wannamakerlipshitzvanderkooywright-00}.   Let $\{y(i)\}_{i\geq 0}$ and $\{\nu(i)\}_{i\geq 0}$  be arbitrary sequences of random variables, and $q(\cdot)$ be the quantization function~(\ref{qfuncdef}).
  When dither is added before quantization, the quantization error sequence, $\{\varepsilon(i)\}_{i\geq 0}$, is
\begin{equation}
\label{defquanterror} \varepsilon(i)=q(y(i)+\nu(i))-(y(i)+\nu(i))
\end{equation}

If the dither sequence, $\{\nu(i)\}_{i\geq 0}$, satisfies the Schuchman conditions, \cite{Schuchman},
then the quantization error sequence, $\{\varepsilon(i)\}_{i\geq
0}$, in~(\ref{defquanterror}) is i.i.d.~uniformly distributed on
$\left[-\Delta/2,\Delta/2\right)$ and independent of the input
sequence $\{y(i)\}_{i\geq 0}$
(see~\cite{SripadSnyder,Gray,Lipshitz}).
A sufficient condition for $\{\nu(i)\}$ to satisfy the Schuchman conditions is for it to be a sequence of i.i.d.~random variables uniformly distributed on $\left[-\Delta/2,\Delta/2\right)$ and independent of the input sequence $\{y(i)\}_{i\geq 0}$. In the sequel, the dither  $\{\nu(i)\}_{i\geq 0}$ satisfies the Schuchman conditions. Hence, the quantization error sequence, $\{\epsilon(i)\}$, is i.i.d.~uniformly distributed on $\left[-\Delta/2,\Delta/2\right)$ and independent of the input sequence $\{y(i)\}_{i\geq 0}$.


 \subsection{Dithered Quantized Consensus With Random Link Failures: Problem Statement}
  \label{ditheredproblemstatement} We now return to the problem formulation of consensus with quantized data with dither added. Introducing the sequence, $\{\nu_{nl}(i)\}_{i\geq
0,1\leq n,l\leq N}$, of i.i.d.~random variables, uniformly
distributed on $\left[-\Delta/2,\Delta/2\right)$, the state update equation for quantized consensus is:
\begin{equation}
\label{randquant}
x_{n}(i+1)=\left(1-\alpha(i)d_{n}(i)\right)x_{n}(i)+
\alpha(i)\sum_{l\in\Omega_{n}(i)}q\left[x_{l}(i)+\nu_{nl}(i)\right],\:\:1\leq
n\leq N
\end{equation}
This equation shows that, before transmitting its state $x_{l}(i)$ to the $n$-th sensor, the sensor $l$ adds the dither
$\nu_{nl}(i)$, then the channel between the sensors~$n$ and~$l$ quantizes this corrupted state, and, finally, sensor~$n$ receives this quantized output. Using eqn.~(\ref{defquanterror}), the state update is
\begin{equation}
\label{randquantupdate}
x_{n}(i+1)=\left(1-\alpha(i)d_{n}\right)x_{n}(i)+
\alpha(i)\sum_{l\in\Omega_{n}(i)}\left[x_{l}(i)+\nu_{nl}(i)
+\varepsilon_{nl}(i)\right]
\end{equation}
The random variables $\nu_{nl}(i)$ are independent of the state $\mathbf{x}(j)$, i.e., the states of all sensors at iteration~$j$, for $j\leq i$. Hence, the collection
$\{\varepsilon_{nl}(i)\}$ consists of i.i.d.~random variables
uniformly distributed on $\left[-\Delta/2,\Delta/2\right)$, and
the random variable $\varepsilon_{nl}(i)$ is also independent of the state
$\mathbf{x}(j)$, $j\leq i$.

We rewrite~(\ref{randquantupdate}) in vector form. Define the random vectors, $\mathbf{\Upsilon}(i)$ and $\mathbf{\Psi}(i)\in\mathbb{R}^{N\times 1}$ with components
\begin{equation}
\label{defUpsilon}
\Upsilon_{n}(i)=-\sum_{l\in\Omega_{n}(i)}\nu_{nl}(i)
\end{equation}
\begin{equation}
\label{defPsi} \Psi_{n}(i) =
-\sum_{l\in\Omega_{n}(i)}\varepsilon_{nl}(i)
\end{equation}
The the $N$~state update equations  in~(\ref{randquantupdate}) become in vector form
\begin{equation}
\label{randquantupdate1}
\mathbf{x}(i+1)=\mathbf{x}(i)-\alpha(i)\left[L(i)\mathbf{x}(i)+\mathbf{\Upsilon}(i)+\mathbf{\Psi}(i)\right]
\end{equation}
where  $\mathbf{\Upsilon}(i)$ and
$\mathbf{\Psi}(i)$ are zero mean vectors, independent of the state $\mathbf{x}(i)$, and have i.i.d. components. Also,
 if~$|\mathcal{M}|$ is the number of realizable network links,
eqns.~(\ref{defUpsilon}) and (\ref{defPsi}) lead to
\begin{equation}
\label{th:QCconv3}
\mathbb{E}\left[\|\mathbf{\Upsilon}(i)\|^{2}\right]=
\mathbb{E}\left[\|\mathbf{\Psi}(i)\|^{2}\right]\leq\frac{|\mathcal{M}|\Delta^{2}}{6},
i\geq 0
\end{equation}

\textbf{Random Link Failures:} We now state the assumption about
the link failure model to be adopted throughout the paper. The
graph Laplacians are
\begin{equation}
\label{Lcond} L(i) = \overline{L}+\widetilde{L}(i),~\forall i\geq
0
\end{equation}
where $\{L(i)\}_{i\geq 0}$ is a sequence of i.i.d.~Laplacian
matrices with mean $\overline{L} = \mathbb{E}\left[L(i)\right]$,
such that $\lambda_{2}\left(\overline{L}\right)>0$ (we just
require the network to be connected on the average.) We do not
make any distributional assumptions on the link failure model.
During the same iteration, the link failures can be spatially
dependent, i.e., correlated across different edges of the network.
This model subsumes the erasure network model, where the link
failures are independent both over space and time. Wireless sensor
networks motivate this model since interference among the sensors
communication correlates the link failures over space, while over
time, it is still reasonable to assume that the channels are
memoryless or independent. We also note that the above assumption
$\lambda_{2}\left(\overline{L}\right)>0$ does not require the
individual random instantiations of $L(i)$ to be connected; in
fact, it is possible to have all the instantiations to be
disconnected. This enables us to capture a broad class of
asynchronous communication models, for example, the random
asynchronous gossip protocol analyzed in~\cite{Boyd-Gossip}
satisfies $\lambda_{2}\left(\overline{L}\right)>0$ and hence falls
under this framework. More generally, in the asynchronous set up,
if the sensors nodes are equipped with independent clocks whose
ticks follow a regular random point process (the ticking instants
do not have an accumulation point, which is true for all renewal
processes, in particular, the Poisson clock
in~\cite{Boyd-Gossip}), and at each tick a random network is
realized with $\lambda_{2}\left(\overline{L}\right)>0$ independent
of the the networks realized in previous ticks (this is the case
with the link formation process assumed in~\cite{Boyd-Gossip}) our
algorithm applies.\footnote{In case the network is static, i.e., the connectivity graph is time-invariant, all the results in the paper apply with $L(i)\equiv \overline{L},~~\forall i$.}

We denote the number of network edges at time $i$ as $M(i)$, where
$M(i)$ is a random subset of the set of all possible edges
$\mathcal{E}$ with $|\mathcal{E}|= N(N-1)/2$. Let $\mathcal{M}$
denote the set of realizable edges. We then have the inclusion
\begin{equation}
\label{lcond123}
M(i)\subset\mathcal{M}\subset\mathcal{E},~~\forall i
\end{equation}
It is important to note that the value of $M(i)$ depends on the
link usage protocol. For example, in the asynchronous gossip
protocol considered in~\cite{Boyd-Gossip}, at each iteration only
one link is active, and hence $M(i)=1$.

\textbf{Independence Assumptions:} We assume that the Laplacian
sequence $\{L(i)\}_{i\geq 0}$ is independent of the dither
sequence $\{\varepsilon_{nl}(i)\}$.

\textbf{Persistence condition:} To obtain convergence, we assume
that the gains $\alpha(i)$ satisfy the following.
\begin{equation}
\label{alphacond} \alpha (i)>0,\:\:\sum_{i\geq 0}\alpha
(i)=\infty,\:\:\sum_{i\geq 0}\alpha^{2}(i)<\infty
\end{equation}
Condition~(\ref{alphacond}) assures that the gains decay to zero, but not too fast. It is standard in stochastic adaptive signal processing and control; it is also used in consensus with noisy communications in~\cite{Mesbahi-Noise,karmoura-randomtopologynoise}.

{\bf Markov property.} Denote the natural filtration of the
process $\mathbf{X}=\left\{\mathbf{x}(i)\right\}_{i\geq0}$ by
$\left\{\mathcal{F}_i^{\mathbf{X}}\right\}_{i\geq 0}$. Because the
dither random variables $\nu_{nl}(i)$, $1\leq n,l\leq N$, are
independent of $\mathcal{F}_i^{\mathbf{X}}$ at any time $i\geq 0$,
and, correspondingly,  the noises $\mathbf{\Upsilon}(i)$ and
$\mathbf{\Psi}(i)$ are independent of $\mathbf{x}(i)$, the process
$\mathbf{X}$ is Markov.





\section{Consensus With Quantized Data: Unbounded Quantized States}
\label{quantizedconsensus}
We consider that the dynamic range of the initial sensor data, whose average we wish to compute, is not known. To avoid
quantizer saturation, the quantizer output takes
values in the countable alphabet~(\ref{Qdef}), and so the channel quantizer has unrestricted dynamic
range. This is the quantizer consensus~(QC) with unbounded range
algorithm. Section~\ref{QCFmajor} studies quantization with unbounded range, i.e., the quantized consensus finite-bit~(QCF) algorithm where the channel quantizers take only a finite number of output
values (finite-bit quantizers).

\label{QCmajor}
 We  comment briefly on the organization of the remaining of this section.
 Subsection~\ref{quantconvproof} proves the a.s.~convergence of the QC~algorithm.
 We characterize the performance of the QC algorithm and derive expressions for
 the mean-squared error in  Subsection~\ref{errorQC}. The tradeoff between m.s.e. and convergence rate
 is studied in Subsection~\ref{ConvRateQC}. Finally,
 we present generalizations to the approach in Subsection~\ref{gen_QC}.

\subsection{QC Algorithm: Convergence} \label{quantconvproof}
We start with the definition of the consensus subspace $\mathcal{C}$ given as
\begin{equation}
\label{ConsSubspace1} \mathcal{C} = \left\{\mathbf{x}\left.\in\mathbb{R}^{N\times
1}\right|\mathbf{x}=a\mathbf{1},~a\in\mathbb{R}\right\}
\end{equation}
We note that any vector $\mathbf{x}\in\mathbb{R}^{N}$ can be uniquely decomposed as
\begin{equation}
\label{decomp}
\mathbf{x}=\mathbf{x}_{\mathcal{C}}+\mathbf{x}_{\mathcal{C}^{\perp}}
\end{equation}
and
\begin{equation}
\label{decomp1}
\left\|\mathbf{x}\right\|^{2}=\left\|\mathbf{x}_{\mathcal{C}}\right\|^{2}+\left\|\mathbf{x}_{\mathcal{C}^{\perp}}\right\|^{2}
\end{equation}
where $\mathbf{x}_{\mathcal{C}}\in\mathcal{C}$ and $\mathbf{x}_{\mathcal{C}^{\perp}}$ belongs to $\mathcal{C}^{\perp}$, the orthogonal subspace of $\mathcal{C}$.
We show that~(\ref{randquantupdate1}), under the model in Subsection~\ref{ditheredproblemstatement}, converges a.s. to a finite point in
$\mathcal{C}$.

Define the component-wise average as
\begin{equation}
\label{compwiseavg}
x_{\mbox{\scriptsize{avg}}}(i)=\frac{1}{N}\mathbf{1}^{T}\mathbf{x}(i)
\end{equation}

We prove the a.s. convergence of the QC~algorithm in two stages.
Theorem~\ref{th:QCconv} proves that the state vector sequence
$\left\{\mathbf{x}(i)\right\}_{i\geq 0}$ converges a.s.~to the
consensus subspace~$\mathcal{C}$.  Theorem~\ref{th:QCavg} then
completes the proof by showing that the sequence  of
component-wise averages,
$\left\{x_{\mbox{\scriptsize{avg}}}(i)\right\}_{i\geq 0}$
converges a.s. to a finite random variable $\theta$. The proof of
Theorem~\ref{th:QCavg} needs a basic result on convergence of
Markov processes and follows the same theme as in~\cite{karmoura-randomtopologynoise}.

{\bf Stochastic approximation: Convergence of Markov processes.}
\label{ConvMarkovProc} We state a slightly modified form , suitable to our needs, of a result
from~\cite{Nevelson}. We start by introducing notation, following~\cite{Nevelson}, see also~\cite{karmoura-randomtopologynoise}.

Let $\mathbf{X}=\left\{\mathbf{x}(i)\right\}_{i\geq 0}$ be Markov in $\mathbb{R}^{N\times 1}$. The generating operator $\mathcal{L}$
is
{\small
\begin{equation}
\label{Lgenop} \mathcal{L}V\left(i,\mathbf{x}\right) =
\mathbb{E}\left[V\left(i+1,\mathbf{x}(i+1)\right)\left|\right.\mathbf{x}(i)=
\mathbf{x}\right]-V\left(i,\mathbf{x}\right)\:\mbox{a.s.}
\end{equation}
}
for functions $V(i,\mathbf{x}),~i\geq
0,~\mathbf{x}\in\mathbb{R}^{N\times 1}$, provided the conditional
expectation exists. We say that $V(i,\mathbf{x})\in
D_{\mathcal{L}}$ in a domain $A$, if $\mathcal{L}V(i,\mathbf{x})$
is finite for all $(i,\mathbf{x})\in A$.

Let the  Euclidean metric be $\rho(\cdot)$. Define the
$\epsilon$-neighborhood of $B\subset\mathbb{R}^{N\times 1}$ and its complementary set
\begin{eqnarray}
\label{Uep} U_{\epsilon}(B) &=& \left\{x\left|\inf_{y\in B}\rho
(x,y)<\epsilon\right.\right\}\\
\label{def_Vep} V_{\epsilon}(B) &=& \mathbb{R}^{N\times 1}\backslash
U_{\epsilon}(B)
\end{eqnarray}
\begin{theorem}[Convergence of Markov Processes]
\label{MarkovProc} Let: $\mathbf{X}$ be a Markov process with
generating operator $\mathcal{L}$; $V(i,\mathbf{x})\in D_{\mathcal{L}}$ a
non-negative function  in the
domain $i\geq 0$, $\mathbf{x}\in\mathbb{R}^{N\times 1}$, and
$B\subset\mathbb{R}^{N\times 1}$. Assume:
%
\begin{eqnarray}
\label{Vcond1}
\hspace{-.5in}
\mbox{\textbf{1) Potential function:}}
\hspace{.5in}
\inf_{i\geq 0,\mathbf{x}\in
V_{\epsilon}(B)}V(i,\mathbf{x})&>&0,\:\:\forall \epsilon> 0\hspace{1in}\\
\label{Vcond2} V(i,\mathbf{x})&\equiv& 0,\:\:\mathbf{x}\in B\\
\label{Vcond3} \lim_{\mathbf{x}\rightarrow B}\sup_{i\geq 0}
V(i,\mathbf{x})&=&0\\
%
%
\label{LVcond1}
\mbox{\textbf{2) Generating operator:}}
\hspace{.895in}
\mathcal{L}V(i,\mathbf{x})&\leq&
g(i)(1+V(i,\mathbf{x}))-\alpha (i)\varphi (i,\mathbf{x})
\hspace{.895in}
\end{eqnarray}
where $\varphi(i,\mathbf{x}),i\geq
0$, $\mathbf{x}\in\mathbb{R}^{N\times 1}$ is a non-negative function
such that
\begin{eqnarray}
\label{LVcond2} \inf_{i,\mathbf{x}\in V_{\epsilon}(B)}\varphi
(i,\mathbf{x})
&>&0,\:\:\forall \epsilon >0\\
%
%
\label{alphacond-1}
\alpha(i) &>&0,\:\:\sum_{i\geq 0}\alpha (i) =
\infty\\
\label{gcond} g(i)&>&0,\:\:\sum_{i\geq 0}g(i)<\infty
\end{eqnarray}
%
Then, the Markov process $\mathbf{X}=\left\{\mathbf{x}(i)\right\}_{i\geq 0}$
with arbitrary initial distribution converges a.s.~to~$B$ as
$i\rightarrow\infty$
\begin{equation}
\label{convmarkov12}
\mathbb{P}\left(\lim_{i\rightarrow\infty}
\rho\left(\mathbf{x}(i),B\right)=0\right)=1
\end{equation}
\end{theorem}
\begin{proof}
For proof, see~\cite{Nevelson,karmoura-randomtopologynoise}.
\end{proof}

\begin{theorem}[{a.s.}~convergence to consensus subspace]
\label{th:QCconv} Consider the quantized distributed averaging
algorithm given in eqns.~(\ref{randquantupdate1}). 
 Then, for arbitrary initial condition,
$\mathbf{x}(0)$, we have
\begin{equation}
\label{th:QCconv1}
\mathbb{P}\left[\lim_{i\rightarrow\infty}\rho(\mathbf{x}(i),\mathcal{C})=0\right]=1
\end{equation}
\end{theorem}
\begin{proof}
The proof uses similar arguments as that of Theorem 3 in~\cite{karmoura-randomtopologynoise}. So we provide the main steps here and only those details which are required for later development of the paper.

The key idea shows that the quantized
iterations satisfy the assumptions of Theorem~\ref{MarkovProc}.
Define the potential function, $V(i,\mathbf{x})$, for the
Markov process $\mathbf{X}$ as
\begin{equation}
\label{th:QCconv4}
V(i,\mathbf{x})=\mathbf{x}^{T}\overline{L}\mathbf{x}
\end{equation}
Then, using the properties of $\overline{L}$  and the continuity
of $V\left(i,\mathbf{x}\right)$,
\begin{equation}
\label{assV1} V(i,\mathbf{x})\equiv
0,~\mathbf{x}\in\mathcal{C}\:\:\mbox{and}\:\: \lim_{\mathbf{x}\rightarrow
\mathcal{C}}\sup_{i\geq 0} V(i,\mathbf{x})=0
\end{equation}
For $\mathbf{x}\in\mathbb{R}^{N\times 1}$, we clearly have
$\rho(\mathbf{x},\mathcal{C})=\left\|\mathbf{x}_{\mathcal{C}^{\perp}}\right\|$.
Using the fact that
$\mathbf{x}^{T}\overline{L}\mathbf{x}\geq\lambda_{2}(\overline{L})\|\mathbf{x}_{\mathcal{C}^{\perp}}\|^{2}$
it then follows
\begin{equation}
\label{assV2.3} \inf_{i\geq 0,\mathbf{x}\in
V_{\epsilon}(\mathcal{C})}V(i,\mathbf{x})\geq\inf_{i\geq
0,\mathbf{x}\in
V_{\epsilon}(\mathcal{C})}\lambda_{2}(\overline{L})\|\mathbf{x}_{\mathcal{C}^{\perp}}\|^{2}\geq
\lambda_{2}\left(\overline{L}\right)\epsilon^{2}
>0
\end{equation}
since $\lambda_{2}\left(\overline{L}\right)>0$. This shows,
together with~(\ref{assV1}), that $V(i,\mathbf{x})$
satisfies~(\ref{Vcond1})--(\ref{Vcond3}).
%
%
%
%

Now consider $\mathcal{L}V(i,\mathbf{x})$. We have using the fact
that
$\widetilde{L}(i)\mathbf{x}=\widetilde{L}(i)\mathbf{x}_{\mathcal{C}^{\perp}}$
and the independence assumptions
\begin{eqnarray}
\label{th:QCconv8} \mathcal{L}V(i,\mathbf{x}) & = &
\mathbb{E}\left[\left(\mathbf{x}(i)-\alpha(i)\overline{L}\mathbf{x}(i)
-\alpha(i)\widetilde{L}(i)\mathbf{x}(i)-\alpha(i)\mathbf{\Upsilon}(i)-\alpha(i)\mathbf{\Psi}(i)\right)^{T}\overline{L}
\left(\mathbf{x}(i)-\alpha(i)\overline{L}\mathbf{x}(i)\phantom{\left(\Upsilon(i)\right)^T}\right.\right.
\nonumber
\\ & & \left.\left.\left.\phantom{\left(\Upsilon(i)\right)^T}
-\alpha(i)\widetilde{L}(i)\mathbf{x}(i)-\alpha(i)\mathbf{\Upsilon}(i)-\alpha(i)\mathbf{\Psi}(i)\right)
\right|\mathbf{x}(i)=\mathbf{x}\right]-\mathbf{x}^{T}\overline{L}\mathbf{x}
\nonumber \\
& \leq & -2\alpha(i)\mathbf{x}^{T}\overline{L}^{2}\mathbf{x}
+\alpha^{2}(i)\lambda_{N}^{3}(\overline{L})
\|\mathbf{x}_{\mathcal{C}^{\perp}}\|^{2}
+\alpha^{2}(i)\lambda_{N}(\overline{L})\mathbb{E}\left[\lambda_{\mbox{\scriptsize{max}}}^{2}
\left(\widetilde{L}(i)\right)\right]\|\mathbf{x}_{\mathcal{C}^{\perp}}\|^{2}\nonumber
\\ & &
+ 2\alpha^{2}(i)\lambda_{N}(\overline{L})
\!\left(\mathbb{E}\left[\|\mathbf{\Upsilon}(i)\|^{2}\right]\right)^{1/2}
\!\left(\mathbb{E}\left[\|\mathbf{\Psi}(i)\|^{2}\right]\right)^{1/2}
+\alpha^{2}(i)\lambda_{N}(\overline{L})\mathbb{E}\left[\|\mathbf{\Upsilon}(i)\|^{2}\right]
\nonumber
\\ & & +\alpha^{2}(i)\lambda_{N}(\overline{L})\mathbb{E}\left[\|\mathbf{\Psi}(i)\|^{2}\right]
\end{eqnarray}
Since $\mathbf{x}^{T}\overline{L}\mathbf{x}\geq\lambda_{2}(\overline{L})\|\mathbf{x}_{\mathcal{C}^{\perp}}\|^{2}$,
the eigenvalues of $\widetilde{L}(i)$ are not greater than $2N$ in
magnitude, and from~(\ref{th:QCconv3})
 get
\begin{eqnarray}
\label{th:QCconv9} \mathcal{L}V(i,\mathbf{x})
& \leq &
-2\alpha(i)\mathbf{x}^{T}\overline{L}^{2}\mathbf{x}+\left(\frac{\alpha^{2}(i)\lambda_{N}^{3}(\overline{L})}{\lambda_{2}(\overline{L})}+\frac{4\alpha^{2}(i)N^{2}\lambda_{N}(\overline{L})}{\lambda_{2}(\overline{L})}\right)\mathbf{x}^{T}\overline{L}\mathbf{x}+\frac{2\alpha^{2}(i)|\mathcal{M}|\Delta^{2}\lambda_{N}(\overline{L})}{3}
\nonumber \\ & \leq &
-\alpha(i)\varphi(i,\mathbf{x})+g(i)\left[1+V(i,\mathbf{x})\right]
\end{eqnarray}
where
\begin{equation}
\label{th:QCconv10}
\varphi(i,\mathbf{x})=2\mathbf{x}^{T}\overline{L}^{2}\mathbf{x},~~g(i)=\alpha^{2}(i)\max\left(\frac{\lambda_{N}^{3}(\overline{L})}{\lambda_{2}(\overline{L})}+\frac{4N^{2}\lambda_{N}(\overline{L})}{\lambda_{2}(\overline{L})},\frac{2|\mathcal{M}|\Delta^{2}\lambda_{N}(L)}{3}\right)
\end{equation}
Clearly, $\mathcal{L}V(i,\mathbf{x})$ and $\varphi(i,\mathbf{x}),g(i)$
 satisfy the remaining assumptions~(\ref{LVcond1})--(\ref{gcond}) of
Theorem~\ref{MarkovProc}; hence,
\begin{equation}
\label{th:QCconv11}
\mathbb{P}\left[\lim_{i\rightarrow\infty}\rho(\mathbf{x}(i),\mathcal{C})=0\right]=1
\end{equation}
\end{proof}
The convergence proof for QC will now be completed in the next Theorem.
%
%
\begin{theorem}[Consensus to finite random variable]
\label{th:QCavg} Consider~(\ref{randquantupdate1}), with arbitrary initial condition
$\mathbf{x}(0)\in\mathbb{R}^{N\times 1}$ and
  the state sequence $\left\{\mathbf{x}(i)\right\}_{i\geq 0}$.
Then, there exists a finite random variable $\theta$ such that
\begin{equation}
\label{th:QCavg100}
\mathbb{P}\left[\lim_{i\rightarrow\infty}\mathbf{x}(i)=\theta\mathbf{1}\right]=1
\end{equation}
\end{theorem}
\begin{proof}
Define the filtration $\left\{\mathcal{F}_{i}\right\}_{i\geq 0}$ as
\begin{equation}
\label{th:QCavg1}
\mathcal{F}_{i}=\mathcal{\sigma}\left\{\mathbf{x}(0),\left\{L(j)\right\}_{0\leq
j<i}, \left\{\mathbf{\Upsilon}(j)\right\}_{0\leq
j<i},\left\{\mathbf{\Psi}(j)\right\}_{0\leq j<i}\right\}
\end{equation}
We will now show that the sequence
$\left\{x_{\mbox{\scriptsize{avg}}}(i)\right\}_{i\geq 0}$ is an
$\mathcal{L}_{2}$-bounded martingale w.r.t.
$\{\mathcal{F}_{i}\}_{i\geq 0}$. In fact,
\begin{equation}
\label{th:QCavg2}
x_{\mbox{\scriptsize{avg}}}(i+1)=x_{\mbox{\scriptsize{avg}}}(i)-\alpha(i)\overline{\Upsilon}(i)-\alpha(i)\overline{\Psi}(i)
\end{equation}
where $\overline{\Upsilon}(i)$ and $\overline{\Psi}(i)$ are the
component-wise averages given by
\begin{equation}
\label{th:QCavg3}
\overline{\Upsilon}(i)=\frac{1}{N}\mathbf{1}^{T}\mathbf{\Upsilon}(i),~~\overline{\Psi}(i)=\frac{1}{N}\mathbf{1}^{T}\mathbf{\Psi}(i)
\end{equation}
Then,
\begin{eqnarray}
\label{th:QCavg4}
\mathbb{E}\left[\left.x_{\mbox{\scriptsize{avg}}}(i+1)\right|\mathcal{F}_{i}\right]
& = &
x_{\mbox{\scriptsize{avg}}}(i)-\alpha(i)\mathbb{E}\left[\left.\overline{\Upsilon}(i)\right|\mathcal{F}_{i}\right]
-\alpha(i)\mathbb{E}\left[\left.\overline{\Psi}(i)\right|\mathcal{F}_{i}\right]\\
\nonumber & = &
x_{\mbox{\scriptsize{avg}}}(i)-\alpha(i)\mathbb{E}\left[\overline{\Upsilon}(i)\right]
-\alpha(i)\mathbb{E}\left[\overline{\Psi}(i)\right]\\
\nonumber & = & x_{\mbox{\scriptsize{avg}}}(i)
\end{eqnarray}
where the last step follows from the fact that
$\mathbf{\Upsilon}(i)$ is independent of $\mathcal{F}_{i}$, and
\begin{eqnarray}
\label{th:QCavg5}
\mathbb{E}\left[\overline{\Psi}(i)\mid\mathcal{F}_{i}\right] & = &
\mathbb{E}\left[\overline{\Psi}(i)\mid\mathbf{x}(i)\right]\\
\nonumber & = & 0
\end{eqnarray}
because $\Psi(i)$ is independent of~$\mathbf{x}(i)$ as argued in Section~\ref{statquant}.

Thus,  the sequence
$\left\{x_{\mbox{\scriptsize{avg}}}(i)\right\}_{i\geq 0}$ is a martingale.
For proving $\mathcal{L}_{2}$ boundedness, note
\begin{eqnarray}
\label{th:QCavg6}
\mathbb{E}\left[x_{\mbox{\scriptsize{avg}}}^{2}(i+1)\right] & = &
\mathbb{E}\left[x_{\mbox{\scriptsize{avg}}}(i)-\alpha(i)\overline{\Upsilon}(i)-\alpha(i)\overline{\Psi}(i)\right]^{2}\\
\nonumber & = &
\mathbb{E}\left[x_{\mbox{\scriptsize{avg}}}^{2}(i)\right]+\alpha^{2}(i)\mathbb{E}\left[\overline{\Upsilon}^{2}(i)\right]+\alpha^{2}(i)\mathbb{E}\left[\overline{\Psi}^{2}(i)\right]+2\alpha^{2}(i)\mathbb{E}\left[\overline{\Upsilon}(i)\overline{\Psi}(i)\right]\\
\nonumber & \leq &
\mathbb{E}\left[x_{\mbox{\scriptsize{avg}}}^{2}(i)\right]+\alpha^{2}(i)\mathbb{E}\left[\overline{\Upsilon}^{2}(i)\right]\\
&&\nonumber
+\alpha^{2}(i)\mathbb{E}\left[\overline{\Psi}^{2}(i)\right]+2\alpha^{2}(i)\left(\mathbb{E}\left[\overline{\Upsilon}^{2}(i)\right]\right)^{1/2}\left(\mathbb{E}\left[\overline{\Psi}^{2}(i)\right]\right)^{1/2}
\end{eqnarray}
Again, it can be shown by using the independence properties and~(\ref{th:QCconv3}) that
\begin{equation}
\label{th:QCavg7}
\mathbb{E}\left[\overline{\Upsilon}^{2}(i)\right] =
\mathbb{E}\left[\overline{\Psi}^{2}(i)\right] \leq
\frac{|\mathcal{M}|\Delta^{2}}{6N^{2}}
\end{equation}
where $M$ is the number of realizable edges in the network
(eqn.~(\ref{lcond123})). It then follows from
eqn.~(\ref{th:QCavg6}) that
\begin{equation}
\label{th:QCavg8}
\mathbb{E}\left[x_{\mbox{\scriptsize{avg}}}^{2}(i+1)\right]\leq\mathbb{E}\left[x_{\mbox{\scriptsize{avg}}}^{2}(i)\right]+\frac{2\alpha^{2}(i)|\mathcal{M}|\Delta^{2}}{3N^{2}}
\end{equation}
Finally, the recursion leads to
\begin{equation}
\label{th:QCavg9}
\mathbb{E}\left[x_{\mbox{\scriptsize{avg}}}^{2}(i)\right]\leq
x_{\mbox{\scriptsize{avg}}}^{2}(0)+\frac{2|\mathcal{M}|\Delta^{2}}{3N^{2}}\sum_{j\geq
0}\alpha^{2}(j)
\end{equation}
Note that in this equation, $x_{\mbox{\scriptsize{avg}}}^{2}(0)$ is bounded since it is the average of the initial conditions, i.e., at time~$0$.
Thus $\left\{x_{\mbox{\scriptsize{avg}}}(i)\right\}_{i\geq 0}$
is an $\mathcal{L}_{2}$-bounded martingale; hence, it converges
a.s. and in $\mathcal{L}_{2}$ to a finite random variable $\theta$
(\cite{Williams}). In other words,
\begin{equation}
\label{th:QCavg90}
\mathbb{P}\left[\lim_{i\rightarrow\infty}x_{\mbox{\scriptsize{avg}}}(i)=\theta\right]=1
\end{equation}
Again, Theorem~\ref{th:QCconv} implies that as
$i\rightarrow\infty$ we have
$\mathbf{x}(i)\rightarrow x_{\mbox{\scriptsize{avg}}}(i)\mathbf{1}$
a.s. This and~(\ref{th:QCavg90}) prove the
Theorem.
\end{proof}


We extend Theorems~\ref{th:QCconv},\ref{th:QCavg} to derive the mean squared (m.s.s.) consensus of the sensor states to the random variable $\theta$ under additional assumptions on the weight sequence $\{\alpha(i)\}_{i\geq 0}$.
\begin{lemma}
\label{corr:QC}
Let the weight sequence $\{\alpha(i)\}_{i\geq 0}$ be of the form:
\begin{equation}
\label{corr:QC100}
\alpha(i)=\frac{a}{(i+1)^{\tau}}
\end{equation}
where $a>0$ and $.5<\tau\leq 1$. Then
the a.s. convergence in Theorem~\ref{th:QCavg} holds in m.s.s. also, i.e.,
\begin{equation}
\label{corr:QC1}
\lim_{i\rightarrow\infty}\mathbb{E}\left[\left(x_{n}(i)-\theta\right)^{2}\right]=0,~~~\forall n
\end{equation}
\end{lemma}
\begin{proof}
The proof is provided in Appendix~\ref{proofs_res}.
\end{proof}

\subsection{QC Algorithm: Mean-Squared Error}
\label{errorQC} Theorem~\ref{th:QCavg} shows that the sensors
reach consensus asymptotically and in fact converge a.s.~to a
finite random variable $\theta$. Viewing $\theta$ as an estimate
of the initial average $r$ (see eqn.~(\ref{def_r})), we
characterize its desirable statistical properties in the following
Lemma.
\begin{lemma}
\label{stat-thetaQC} Let $\theta$ be as given in
Theorem~\ref{th:QCavg} and $r$, the initial average, as given in
eqn.~(\ref{def_r}). Define
\begin{equation}
\label{mserror-1} \zeta = \mathbb{E}\left[\mathbf{\theta} -
r\right]^{2}
\end{equation}
to be the m.s.e. Then, we have:
\begin{itemize}[\setlabelwidth{1)}]
\item{\textbf{1)}} \textbf{Unbiasedness}:\hspace{1.5in}  $\mathbb{E}\left[\theta\right]=r$
%
\item{\textbf{2)}} \textbf{M.S.E. Bound}:\hspace{1.45in} $\zeta
\leq \frac{2|\mathcal{M}|\Delta^{2}}{3N^{2}}\sum_{j\geq
0}\alpha^{2}\left(j\right)$
%
\end{itemize}
\end{lemma}
\begin{proof}
The proof follows from the arguments presented in the proof of
Theorem~\ref{th:QCavg} and is omitted.
\end{proof}
We note that the m.s.e. bound in Lemma~\ref{stat-thetaQC} is
conservative. Recalling the definition of $M(i)$, as the number of
active links at time $i$ (see eqn.~(\ref{lcond123})), we have (by
revisiting the arguments in the proof of Theorem~\ref{th:QCavg})
\begin{equation}
\label{stat-thetaint1} \zeta \leq
\frac{2\Delta^{2}}{3N^{2}}\sum_{j\geq
0}\alpha^{2}\left(j\right)\mathbb{E}\left[|M(i)|^{2}\right]
\end{equation}
(Note that the term $\sum_{j\geq
0}\alpha^{2}\left(j\right)\mathbb{E}\left[|M(i)|^{2}\right]$ is
well-defined as
$\mathbb{E}\left[|M(i)|^{2}\right]\leq|\mathcal{M}|^{2},~\forall
i$.) In case, we have a fixed (non-random) topology,
$M(i)=\mathcal{M},~\forall i$ and the bound in
eqn.~(\ref{stat-thetaint1}) reduces to the one in
Lemma~\ref{stat-thetaQC}. For the asynchronous gossip protocol
in~\cite{Boyd-Gossip}, $|M(i)|=1,~\forall i$, and hence
\begin{equation}
\label{stat-thetaint2} \zeta_{\mbox{\scriptsize{gossip}}} \leq
\frac{2\Delta^{2}}{3N^{2}}\sum_{j\geq 0}\alpha^{2}\left(j\right)
\end{equation}

Lemma~\ref{stat-thetaQC} shows that, for a given $\Delta$, $\zeta$
can be made arbitrarily small by properly scaling the weight
sequence, $\{\alpha(i)\}_{i\geq 0}$. We formalize this.
Given an arbitrary weight sequence, $\{\alpha(i)\}_{i\geq 0}$,
which satisfies the persistence condition~(\ref{alphacond}),
define the scaled weight sequence, $\{\alpha_{s}(i)\}_{i\geq 0}$,
as
\begin{equation}
\label{weights} \alpha_{s}(i)=s\alpha(i),~\forall i\geq 0
\end{equation}
where, $s>0$, is a constant scaling factor. Clearly, such a scaled
weight sequence satisfies the persistence condition~(\ref{alphacond}), and the
m.s.e. $\zeta_{s}$ obtained by using this scaled weight sequence
is given by
\begin{equation}
\label{msbound-2s} \zeta_{s} \leq
\frac{2|\mathcal{M}|\Delta^{2}s^2}{3N^{2}}\sum_{j\geq
0}\alpha^{2}\left(j\right)
\end{equation}
showing that, by proper scaling of the weight
sequence, the m.s.e.~can be made arbitrarily small.

However, reducing the
m.s.e. by scaling the weights in this way will reduce the convergence
rate of the algorithm. This tradeoff is considered in the
next subsection.

\subsection{QC Algorithm: Convergence Rate}
\label{ConvRateQC} A detailed pathwise convergence rate analysis can be carried out for the QC algorithm using strong approximations like laws of iterated logarithms etc., as is the case with a large class of stochastic approximation algorithms. More generally, we can study formally some moderate deviations asymptotics (\cite{Dupuis-KushnerLDPstochapp},\cite{Kushner-Yin}) or take recourse to concentration inequalities (\cite{Borkar-stochapp}) to characterize convergence rate. Due to space limitations we
do not pursue such analysis in this paper; rather, we present convergence rate analysis for the state sequence $\{\mathbf{x}(i)\}_{i\geq 0}$ in the m.s.s. and that of the mean state vector
sequence. We start by studying the convergence of the mean state vectors, which is simple, yet illustrates an interesting trade-off between the achievable convergence rate and the mean-squared error $\zeta$ through design of the weight sequence $\{\alpha(i)\}_{i\geq 0}$.

From the asymptotic unbiasedness of $\theta$ we have
\begin{equation}
\label{cr1}
\lim_{i\rightarrow\infty}\mathbb{E}\left[\mathbf{x}(i)\right]=r\mathbf{1}
\end{equation}
Our objective is to determine the rate at which the sequence
$\{\mathbb{E}\left[\mathbf{x}(i)\right]\}_{i\geq 0}$ converges to
$r\mathbf{1}$.

\begin{lemma}
\label{convrQC} Without loss of generality, make the
assumption
\begin{equation}
\label{cr3}
\alpha(i)\leq\frac{2}{\lambda_{2}\left(\overline{L}\right)+\lambda_{N}(\overline{L})},~\forall
i
\end{equation}
(We note that this holds eventually, as the $\alpha(i)$ decrease
to zero.) Then,
\begin{equation}
\label{cr5}
\left\|\mathbb{E}\left[\mathbf{x}(i)\right]-r\mathbf{1}\right\|\leq\left(e^{-\lambda_{2}\left(\overline{L}\right)\left(\sum_{0\leq
j\leq
i-1}\alpha(j)\right)}\right)\left\|\mathbb{E}\left[\mathbf{x}(0)\right]-r\mathbf{1}\right\|
\end{equation}
\end{lemma}
\begin{proof}
We note that the mean state propagates as
\begin{equation}
\label{meanprop}
\mathbb{E}\left[\mathbf{x}(i+1)\right]=\left(I-\alpha(i)\overline{L}\right)\mathbb{E}\left[\mathbf{x}(i)\right],~\forall
i
\end{equation}
The proof then follows from~\cite{karmoura-randomtopologynoise}
and is omitted.
\end{proof}
It follows from Lemma~\ref{convrQC} that the rate at which the
sequence $\{\mathbb{E}\left[\mathbf{x}(i)\right]\}_{i\geq 0}$
converges to $r\mathbf{1}$ is closely related to the rate at which
the weight sequence, $\alpha(i)$, sums to infinity. On the other
hand, to achieve a small bound $\zeta$ on the m.s.e, see
lemma~\ref{mserror-1} in Subsection~\ref{errorQC}, we need to make
the weights small, which reduces the convergence rate of the
algorithm. The parameter $s$ introduced in eqn.~(\ref{weights})
can then be viewed as a scalar control parameter, which can be
used to trade-off between precision (m.s.e.) and convergence rate.
More specifically, if a cost structure is imposed on the consensus
problem, where the objective is a function of the m.s.e. and the
convergence rate, one may obtain the optimal scaling $s$
minimizing the cost from the pareto-optimal curve generated by
varying $s$. This is significant, because the algorithm allows one
to trade off m.s.e. vs. convergence rate, and in particular, if
the application requires precision (low m.s.e.), one can make the
m.s.e. arbitrarily small irrespective of the quantization
step-size $\Delta$. It is important to note in this context, that
though the algorithms in~\cite{Tuncer,Kashyap} lead to finite
m.s.e., the resulting m.s.e. is proportional to $\Delta^{2}$,
which may become large if the step-size $\Delta$ is chosen to be
large.

Note that this tradeoff is established between the convergence
rate of the mean state vectors and the m.s.e.~of the limiting
consensus variable~$\theta$. But, in general, even for more
appropriate measures of the convergence rate, we expect that,
intuitively, the same tradeoff will be exhibited, in the sense
that the rate of convergence will be closely related to the rate
at which the weight sequence, $\alpha(i)$, sums to infinity. We end this subsection by studying the m.s.s. convergence rate of the state sequence $\{\mathbf{x}(i)\}_{i\geq 0}$ which is shown to exhibit a similar trade-off.
\begin{lemma}
\label{QC:convmss}
Let the weight sequence $\{\alpha(i)\}_{i\geq 0}$ be of the form:
\begin{equation}
\label{QC:convmss1}
\alpha(i)=\frac{a}{(i+1)^{\tau}}
\end{equation}
where $a>0$ and $.5<\tau\leq 1$. Then the m.s.s. error evolves as follows:
For every $0<\varepsilon<\frac{2\lambda_{2}^{2}(\overline{L})}{\lambda_{N}(L)}$, there exists $i_{\varepsilon}\geq 0$, such that, for all $i\geq i_{\varepsilon}$ we have
\begin{eqnarray}
\label{QC:convmss2}
\mathbb{E}\left[\left\|\mathbf{x}(i)-r\mathbf{1}\right\|^{2}\right] & \leq & \frac{1}{\lambda_{2}(\overline{L})}e^{-\left(2
\frac{\lambda_{2}^{2}(\overline{L})}{\lambda_{N}(\overline{L})}
-\varepsilon\right)\sum_{j=i_{\varepsilon}}^{i-1}\alpha(j)}
\mathbb{E}\left[\left\|\mathbf{x}_{\mathcal{C}^{\perp}}(i_{\varepsilon})\right\|^{2}\right]\\
&+& \frac{1}{\lambda_{2}(\overline{L})}\sum_{j=i_{\varepsilon}}^{i-1}\left[\left(e^{-\left(2
\frac{\lambda_{2}^{2}(\overline{L})}{\lambda_{N}(\overline{L})}-\varepsilon\right)
\sum_{l=j+1}^{i-1}\alpha(l)} \right)g(j)\right]\nonumber 
+\frac{2|\mathcal{M}|\Delta^{2}}{3}\sum_{j=0}^{i-1}\alpha^{2}(j)
\end{eqnarray}
\end{lemma}
\begin{proof}
The proof is provided in Appendix~\ref{proofs_res}.
\end{proof}
From the above we note that slowing up the sequence $\{\alpha(i)\}_{i\geq 0}$ decreases the polynomial terms on the R.H.S. of eqn.~(\ref{QC:convmss2}), but increases the exponential terms and since the effect of exponentials dominate that of the polynomials we see a similar trade-off between m.s.e. and convergence rate (m.s.s.) as observed when studying the mean state vector sequence above.

\subsection{QC Algorithm: Generalizations}
\label{gen_QC} 
 The QC~algorithm can be extended to handle more complex situations of
imperfect communication. For instance, we may incorporate
Markovian link failures (as
in~\cite{karmoura-randomtopologynoise}) and time-varying
quantization step-size with the same type of analysis.

Markovian packet dropouts can be an issue in some practical wireless sensor network scenarios, where random environmental phenomena like scattering may lead to temporal dependence in the link quality. Another situation arises in networks of mobile agents, where physical aspects of the transmission like channel coherence time, channel fading effects are related to the mobility of the dynamic network. A general analysis of all such scenarios is beyond the scope of the current paper. However, when temporal dependence is manifested through a state dependent Laplacian (this occurs in mobile networks, formation control problems in multi-vehicle systems), under fairly general conditions, the link quality can be modeled as a temporal Markov process as in~\cite{karmoura-randomtopologynoise} (see Assumption~\textbf{1.2} in~\cite{karmoura-randomtopologynoise}.) Due to space limitations of the current paper, we do not present a detailed analysis in this context and refer the interested reader to~\cite{karmoura-randomtopologynoise}, where such temporally Markov link failures were addressed in detail, though in the context of unquantized analog transmission.

The current paper focuses on quantized transmission of data and neglects the effect of additive analog noise. Even in such a situation of digital transmission, the message decoding process at the receiver may lead to analog noise. Our approach can take into account such generalized distortions and the main results will continue to hold. For analysis purposes, temporally independent zero mean analog noise can be incorporated as an additional term on the R.H.S. of eqn.~(\ref{randquantupdate1}) and subsequently absorbed into the zero mean vectors $\mathbf{\Psi}(i),\mathbf{\Upsilon}(i)$. Digital transmission where bits can get flipped due to noise would be more challenging to address.

The case
of time-varying quantization may be relevant in many practical
communication networks, where because of a bit-budget, as time
progresses the quantization may become coarser (the step-size
increases). It may also arise if one considers a rate allocation
protocol with vanishing rates as time progresses
(see~\cite{Scaglione-ICASSP08}). In that case, the quantization
step-size sequence, $\{\Delta(i)\}_{i\geq 0}$ is time-varying with
possibly
\begin{equation}
\label{gen_QC1} \limsup_{i\rightarrow\infty}\Delta(i)=\infty
\end{equation}
Also, as suggested in~\cite{Tuncer-dither}, one may consider a
rate allocation scheme, in which the quantizer becomes finer as
time progresses. In that way, the quantization step-size sequence,
$\{\Delta(i)\}_{i\geq 0}$ may be a decreasing sequence.

Generally, in a situation like this to attain consensus the link
weight sequence $\{\alpha(i)\}_{i\geq 0}$ needs to satisfy a
generalized persistence condition of the form
\begin{equation}
\label{gen_QC2} \sum_{i\geq 0}\alpha(i)=\infty,~~\sum_{i\geq
0}\alpha^{2}(i)\Delta^{2}(i)<\infty
\end{equation}
Note, when the quantization step-size is bounded, this reduces to
the persistence condition assumed earlier. We state without proof
the following result for time-varying quantization case.
\begin{theorem}
\label{thgen_QC} Consider the QC algorithm with time-varying
quantization step size sequence $\{\Delta(i)\}_{i\geq 0}$ and let
the link weight sequence $\{\alpha(i)\}_{i\geq 0}$ satisfy the
generalized persistence condition in eqn.~(\ref{gen_QC2}). Then
the sensors reach consensus to an a.s. finite random variable. In
other words, there exists an a.s. finite random variable $\theta$,
such that,
\begin{equation}
\label{gen_QC3}
\mathbb{P}\left[\lim_{i\rightarrow\infty}x_{n}(i)=\theta,~\forall
n\right]=1
\end{equation}
Also, if $r$ is the initial average, then
\begin{equation}
\label{gen_QC4}
\mathbb{E}\left[\left(\theta-r\right)^{2}\right]\leq\frac{2|\mathcal{M}|}{3N^{2}}\sum_{i\geq
0}\alpha^{2}(i)\Delta^{2}(i)
\end{equation}
\end{theorem}
It is clear that in this case also, we can trade-off m.s.e. with
convergence rate by tuning a scalar gain parameter $s$ associated
with the link weight sequence.

\section{Consensus with Quantized Data: Bounded Initial Sensor State}
\label{QCFmajor}
We consider consensus with
quantized data and bounded range quantizers when the initial sensor states are
bounded, and this bound is known \emph{a priori}. We
show that finite bit quantizers (whose outputs take
only a finite number of values) suffice.
 The algorithm QCF that we consider is a simple modification of the
QC algorithm of Section~\ref{quantizedconsensus}. The good performance of the QCF algorithm relies
on the fact that, if the initial sensor states are
bounded, the state sequence, $\left\{\mathbf{x}(i)\right\}_{i\geq 0}$
generated by the QC algorithm remains uniformly bounded with high
probability, as we prove here. In this case, channel quantizers with finite
dynamic range perform well with high probability.

We briefly state  the QCF problem in Subsection~\ref{ProbFormQCF}. Then, Subsection~\ref{samppathbounds} shows that with high probability
 the sample paths generated by the QC algorithm are uniformly bounded, when the initial sensor states are
bounded. Subsection~\ref{QCF} proves that QCF achieves asymptotic consensus. Finally, Subsections~\ref{statpropQCF} and~\ref{NumStudQCF} analyze its statistical properties, performance, and tradeoffs.

\subsection{QCF Algorithm: Statement}
\label{ProbFormQCF}
The QCF~algorithm modifies the QC~algorithm by restricting the alphabet of the quantizer to be finite. It assumes that the initial sensor state $\mathbf{x}(0)$, whose average we wish to compute, is known to be bounded. Of course, even if the initial state is bounded, the states of QC can become unbounded. The good performance of QCF is a consequence of the fact that, as our analysis will show, the states $\left\{\mathbf{x}(i)\right\}_{i\geq 0}$ generated by the QC algorithm when started with a bounded initial state $\mathbf{x}(0)$ remain uniformly bounded with high probability.

 The following are the assumptions underlying QCF. We let the the state sequence for QCF be represented by
$\left\{\widetilde{\mathbf{x}}(i)\right\}_{i\geq 0}$.
 \begin{enumerate}
 \item
 \label{boundedinitialstate} Bounded initial state. Let $b>0$. The QCF initial state $\widetilde{\mathbf{x}}(0)=x_{n}(0)$ is bounded to the  set~$\mathcal{B}$ known \`a priori
\begin{equation}
\label{mathcalB1} \mathcal{B}=\left\{\mathbf{y}\in\mathbb{R}^{N\times
1}~|~|y_{n}|\leq b< +\infty\right\}
\end{equation}
\item
\label{uniformquantizer}
 Uniform quantizers and finite alphabet.
Each inter-sensor communication
channel in the network uses a uniform
$\lceil\log_{2}(2p+1)\rceil$ bit quantizer with step-size
$\Delta$, where $p>0$ is an integer. In other words, the quantizer
output takes only $2p+1$ values, and the quantization alphabet is
given by
\begin{equation}
\label{QCFbound1} \widetilde{\mathcal{Q}}=\{l\Delta~|~l=0,\pm
1,\cdots ,\pm p\}
\end{equation}
Clearly, such a quantizer will not saturate if the input falls in the range
$\left[(-p-1/2)\Delta,(p+1/2)\Delta\right)$; if the input goes out of that range, the quantizer saturates.
\item
\label{uniformiidnoise}
 Uniform i.i.d.~noise. Like with
QC, the $\left\{\nu_{nl}(i)\right\}_{i\geq 0,1\leq n,l\leq N}$ are
a sequence of i.i.d.~random variables uniformly distributed on
$\left[-\Delta/2,\Delta/2\right)$.

\item \label{linkfailuremodel} The link failure model is the same
as used in QC.
\end{enumerate}
Given this setup, we
 present the distributed QCF algorithm,
assuming that the sensor network is connected.
%
%
%
 The state sequence,
$\left\{\widetilde{\mathbf{x}}(i)\right\}_{i\geq 0}$ is given by the following Algorithm.
\begin{algorithm}
\textbf{Initialize}

 $\widetilde{x}_{n}(0)= x_{n}(0),\:\forall n$\; $i=0$\;\: \Begin{ \While {$\sup_{1\leq n\leq
N}\sup_{l\in\Omega_{n}(i)}|(\widetilde{x}_{l}(i)
+\nu_{nl}(i))|<(p+1/2)\Delta$}{\;
$\widetilde{x}_{n}(i+1)=(1-\alpha(i)d_{n}(i))\widetilde{x}_{n}(i)+
\alpha(i)\sum_{l\in\Omega_{n}(i)}q(\widetilde{x}_{l}(i)+\nu_{nl}(i)),
\:\forall n$\; $i=i+1$\;}} Stop the algorithm and reset all the
sensor states to zero \caption{QCF\label{QCF-alg}}
\end{algorithm}

The last step of the algorithm can be distributed, since
the network is connected.

\subsection{Probability Bounds on Uniform Boundedness of
Sample Paths of QC} \label{samppathbounds} The analysis of the QCF
algorithm requires uniformity properties of the sample paths
generated by the QC algorithm. This is necessary, because the QCF
algorithm follows the QC algorithm till one of the quantizers gets
overloaded. The uniformity properties require establishing
statistical properties of the supremum taken over the sample paths,
which is carried out in this subsection. We show that the state
vector sequence, $\left\{\mathbf{x}(i)\right\}_{i\geq 0}$,
generated by the QC algorithm is uniformly bounded with high
probability. The proof follows by splitting the sequence
$\{\mathbf{x}(i)\}_{i\geq 0}$ as the sum of the sequences
$\{\mathbf{x}_{\mbox{\scriptsize{avg}}}(i)\}_{i\geq 0}$ and
$\{\mathbf{x}_{\mathcal{C}^{\perp}}(i)\}_{i\geq 0}$ for which we
establish uniformity results. The proof is lengthy and uses mainly
maximal inequalities for submartingale and supermartingale
sequences.

Recall that the state vector at any time $i$ can be decomposed
orthogonally as
\begin{equation}
\label{eq:sbound1}
\mathbf{x}(i)=x_{\mbox{\scriptsize{avg}}}(i)\mathbf{1}+\mathbf{x}_{\mathcal{C}^{\perp}}(i)
\end{equation}
where the consensus subspace, $\mathcal{C}$, is given in
eqn.~(\ref{ConsSubspace1}). We provide probability bounds on the
sequences $\left\{x_{\mbox{\scriptsize{avg}}}(i)\right\}_{i\geq 0}$ and
$\left\{\mathbf{x}_{\mathcal{C}^{\perp}}(i)\right\}_{i\geq 0}$ and then use
an union bound to get the final result.

The rest of the subsection concerns the proof of Theorem~\ref{th:sampbound} which involves several intermediate lemmas as stated below, whose proofs are provided in Appendix~\ref{app:sampbound}.


We need the following result.
\begin{lemma}
\label{Wsupmart} Consider the QC algorithm stated in
Section~\ref{randomtopology} and let $\{\mathbf{x}(i)\}_{i\geq 0}$
be the state it sequence generates. Define the function
$W(i,\mathbf{x}),i\geq 0,~\mathbf{x}\in\mathbb{R}^{N\times 1}$, as
\begin{equation}
\label{Wdef}
W(i,\mathbf{x})=\left(1+V\left(i,\mathbf{x}\right)\right)\prod_{j\geq
i}\left[1+g(j)\right]
\end{equation}
where $V(i,\mathbf{x})=\mathbf{x}^{T}\overline{L}\mathbf{x}$ and
$\{g(j)\}_{j\geq 0}$ is defined in
eqn.~(\ref{th:QCconv10}).\footnote{The above function is
well-defined because the term $\prod_{j\geq i}\left[1+g(j)\right]$
is finite for any $j$, by the persistence condition on the weight
sequence.} Then, the process $\{W(i,\mathbf{x}(i)\}_{i\geq 0}$ is
a non-negative supermartingale with respect to the filtration
$\{\mathcal{F}_{i}\}_{i\geq 0}$ defined in eqn.~(\ref{th:QCavg1}).
\end{lemma}

The next Lemma bounds the sequence
$\left\{\mathbf{x}_{\mathcal{C}^{\perp}}(i)\right\}_{i\geq 0}$.

\begin{lemma}
\label{lm:sbound1} Let $\left\{\mathbf{x}(i)\right\}_{i\geq 0}$ be
the state vector sequence generated by the QC algorithm, with the
initial state $\mathbf{x}(0)\in\mathbb{R}^{N\times 1}$. Consider
the orthogonal decomposition:
\begin{equation}
\label{lm:sbound2}
\mathbf{x}(i)=x_{\mbox{\scriptsize{avg}}}(i)\mathbf{1}+\mathbf{x}_{\mathcal{C}^{\perp}}(i),~\forall
i
\end{equation}
 Then, for any $a>0$, we have
\begin{equation}
\label{lm:sbound3} \mathbb{P}\left[\sup_{j\geq
0}\|\mathbf{x}_{\mathcal{C}^{\perp}}(j)\|^{2}>a\right]\leq
\frac{\left(1+\mathbf{x}(0)^{T}\overline{L}\mathbf{x}(0)\right)\prod_{j\geq
0}(1+g(j))}{1+a\lambda_{2}(\overline{L})}
\end{equation}
where $\{g(j)\}_{j\geq 0}$ is defined in eqn.~(\ref{th:QCconv10}).
\end{lemma}

Next, we provide probability bounds on the uniform boundedness of
$\left\{x_{\mbox{\scriptsize{avg}}}(i)\right\}_{i\geq 0}$.

\begin{lemma}
\label{lm:sbound5} Let $\left\{x_{\mbox{\scriptsize{avg}}}(i)\right\}_{i\geq
0}$ be the average sequence generated by the QC algorithm, with an
initial state $\mathbf{x}(0)\in\mathbb{R}^{N\times 1}$. Then, for
any $a>0$,
\begin{equation}
\label{lm:sbound6} \mathbb{P}\left[\sup_{j\geq
0}|x_{\mbox{\scriptsize{avg}}}(j)|> a\right]\leq
\frac{\left[x_{\mbox{\scriptsize{avg}}}^{2}(0)+\frac{2|\mathcal{M}|\Delta^{2}}{3N^{2}}\sum_{j\geq
0}\alpha^{2}(j)\right]^{1/2}}{a}
\end{equation}
\end{lemma}

\begin{theorem}
\label{th:sampbound} Let $\left\{\mathbf{x}(i)\right\}_{i\geq 0}$ be the
state vector sequence generated by the QC algorithm, with an
initial state $\mathbf{x}(0)\in\mathbb{R}^{N\times 1}$. Then, for
any $a>0$,
\begin{equation}
\label{th:sampbound1} \mathbb{P}\left[\sup_{j\geq
0}\left\|\mathbf{x}(j)\right\|>a\right]\leq
\frac{\left[2Nx_{\mbox{\scriptsize{avg}}}^{2}(0)+\frac{4|\mathcal{M}|\Delta^{2}}{3N}\sum_{j\geq
0}\alpha^{2}(j)\right]^{1/2}}{a}+\frac{\left(1+\mathbf{x}(0)^{T}\overline{L}\mathbf{x}(0)\right)\prod_{j\geq
0}(1+g(j))}{1+\frac{a^{2}}{2}\lambda_{2}(\overline{L})}
\end{equation}
where $\{g(j)\}_{j\geq 0}$ is defined in eqn.~(\ref{th:QCconv10}).
\end{theorem}

We now state as a Corollary the result on the
boundedness of the sensor states, which will be used in analyzing
the performance of the QCF algorithm.
\begin{corollary}
\label{corr:sampbound} Assume that the initial sensor state,
$\mathbf{x}(0)\in\mathcal{B}$, where $\mathcal{B}$ is given in
eqn.~(\ref{mathcalB1}). Then, if $\left\{\mathbf{x}(i)\right\}_{i\geq 0}$ is
the state sequence generated by the QC algorithm starting from
the initial state, $\mathbf{x}(0)$, we have, for any $a>0$,
\begin{equation}
\label{corr:sampbound1}
 \mathbb{P}\left[\sup_{1\leq n\leq N,j\geq
0}|x_{n}(j)|>a\right] \leq
\frac{\left[2Nb^{2}+\frac{4|\mathcal{M}|\Delta^{2}}{3N}\sum_{j\geq
0}\alpha^{2}(j)\right]^{1/2}}{a}+\frac{\left(1+N\lambda_{N}(\overline{L})b^{2}\right)\prod_{j\geq
0}(1+g(j))}{1+\frac{a^{2}}{2}\lambda_{2}(\overline{L})}
\end{equation}
where $\{g(j)\}_{j\geq 0}$ is defined in eqn.~(\ref{th:QCconv10}).
\end{corollary}

\subsection{Algorithm QCF: Asymptotic Consensus}
\label{QCF}
We show that the QCF algorithm, given in Subsection~\ref{ProbFormQCF}, converges a.s. to a finite random variable and the sensors
reach consensus asymptotically.
\begin{theorem}[QCF: {a.s.}~asymptotic consensus]
\label{lm:QCF} Let $\left\{\widetilde{\mathbf{x}}(i)\right\}_{i\geq 0}$ be
the state vector sequence generated by the QCF algorithm, starting
from an initial state
$\widetilde{\mathbf{x}}(0)=\mathbf{x}(0)\in\mathcal{B}$. Then, the
sensors reach consensus asymptotically a.s. In other words, there
exists an a.s. finite random variable $\widetilde{\theta}$ such
that
\begin{equation}
\label{lm:QCF1}
\mathbb{P}\left[\lim_{i\rightarrow\infty}\widetilde{\mathbf{x}}(i)=\widetilde{\theta}\mathbf{1}\right]=1
\end{equation}
\end{theorem}

\begin{proof}
For the proof, consider the sequence $\left\{\mathbf{x}(i)\right\}_{i\geq 0}$
generated by the QC algorithm, with the same initial state
$\mathbf{x}(0)$. Let $\theta$ be the a.s. finite random variable
(see eqn.~\ref{th:QCconv11}) such that
\begin{equation}
\label{lm:QCF2}
\mathbb{P}\left[\lim_{i\rightarrow\infty}\mathbf{x}(i)=\theta\mathbf{1}\right]=1
\end{equation}
It is clear that
\begin{equation}
\label{lm:QCF3} \widetilde{\theta} = \left\{\begin{array}{ll}
                    \theta & \mbox{on $\left\{\sup_{i\geq 0}\sup_{1\leq
n\leq
N}\sup_{l\in\Omega_{n}(i)}|x_{l}(i)+\nu_{nl}(i)|<(p+\frac{1}{2})\Delta\right\}$} \\
                    0 & \mbox{otherwise}
                   \end{array}
          \right.
\end{equation}
In other words, we have
\begin{equation}
\label{lm:QCF4}
\widetilde{\theta}=\theta\mathbb{I}\left(\sup_{i\geq 0}\sup_{1\leq
n\leq
N}\sup_{l\in\Omega_{n}(i)}|x_{l}(i)+\nu_{nl}(i)|<(p+\frac{1}{2})\Delta\right)
\end{equation}
where $\mathbb{I}(\cdot)$ is the indicator function. Since
$\left\{\sup_{i\geq 0}\sup_{1\leq n\leq
N}\sup_{l\in\Omega_{n}(i)}|x_{l}(i)+\nu_{nl}(i)|<(p+1/2)\Delta\right\}$
is a measurable set, it follows that $\widetilde{\theta}$ is a
random variable.
\end{proof}

\subsection{QCF: $\epsilon$-Consensus}
\label{statpropQCF}
 Recall the QCF algorithm in Subsection~\ref{ProbFormQCF} and the assumptions~\ref{boundedinitialstate})--\ref{linkfailuremodel}). A key step
is that, if we run the QC algorithm using finite bit quantizers
with finite alphabet $\widetilde{\mathcal{Q}}$ as in
eqn.~(\ref{QCFbound1}), the only way for an error to occur is for
one of the quantizers to saturate. This is the
intuition behind the design of the QCF algorithm.

Theorem~\ref{lm:QCF} shows that the QCF sensor states
asymptotically reach consensus, converging a.s.~to a finite random
variable $\widetilde{\theta}$. The next series of results address
the question of how close is this consensus to the desired
average~$r$ in~(\ref{def_r}).  Clearly, this depends on the QCF
design:
\begin{inparaenum}[1)] \item the quantizer parameters (like the number of levels $2p+1$ or the quantization step~$\Delta$); \item the random network topology ;  and \item the gains~$\alpha$.
\end{inparaenum}

We define the following performance metrics which characterize the
performance of the QCF algorithm.
\begin{definition}[Probability of $\epsilon$-consensus and consensus-consistent]
 \label{def:epsilonconsensus} The probability
 of $\epsilon$-consensus is defined as
\begin{equation}
\label{eq:epsilonconsensus}
T(G,b,\mathbf{\alpha},\epsilon,p,\Delta) =
\mathbb{P}\left[\lim_{i\rightarrow\infty}\sup_{1\leq n\leq
N}|\widetilde{x}_{n}(i)-r|<\epsilon\right]
\end{equation}
Note that the argument $G$ in the definition of $T(\cdot)$
emphasizes the influence of the network configuration, whereas $b$
is given in eqn.~(\ref{mathcalB1}).

The QCF algorithm is
consensus-consistent\footnote{Consensus-consistent means for
arbitrary $\epsilon>0$, the QCF quantizers can be designed so that
the QCF states get within an $\epsilon$-ball of~$r$ with arbitrary
high probability. Thus, a consensus-consistent algorithm trades
off accuracy with bit-rate.} \emph{iff} for every $G,b,\epsilon>0$
and $0<\delta<1$, there exists quantizer parameters $p,\Delta$ and
weights $\{\alpha(i)\}_{i\geq 0}$, such that
\begin{equation}
\label{eq:epsilonconsensus1}
T(G,b,\mathbf{\alpha},\epsilon,p,\Delta)>1-\delta
\end{equation}
\end{definition}
Theorem~\ref{th:QCF} characterizes the probability of
$\epsilon$-consensus, while Proposition~\ref{prop_T} considers
several tradeoffs between the probability of achieving consensus
and the quantizer parameters and network topology, and, in
particular, shows that the QCF algorithm is consensus-consistent. We need the following Lemma to prove Theorem~\ref{th:QCF}.

\begin{lemma}
\label{lm:QCF5} Let $\widetilde{\theta}$ be defined as in
Theorem~\ref{lm:QCF}, with the initial state
$\widetilde{\mathbf{x}}(0)=\mathbf{x}(0)\in\mathcal{B}$. The
 desired average, $r$, is given in~(\ref{def_r}).
 Then, for any $\epsilon >0$, we have
\begin{eqnarray}
\label{lm:QCF7}
\mathbb{P}\left[|\widetilde{\theta}-r|\geq\epsilon\right]&\leq&
\frac{2|\mathcal{M}|\Delta^{2}}{3N^{2}\epsilon^{2}}\sum_{j\geq
0}\alpha^{2}(j)+\frac{\left[2Nb^{2}+\frac{4|\mathcal{M}|\Delta^{2}}{3N}\sum_{j\geq
0}\alpha^{2}(j)\right]^{1/2}}{p\Delta}\\
&&+\frac{\left(1+N\lambda_{N}(\overline{L})b^{2}\right)\prod_{j\geq
0}(1+g(j))}{1+\frac{p^{2}\Delta^{2}}{2}\lambda_{2}(\overline{L})}
\end{eqnarray}
where $\{g(j)\}_{j\geq 0}$ is defined in eqn.~(\ref{th:QCconv10}).
\end{lemma}

\begin{proof}
The proof is provided in Appendix~\ref{app:3}.
\end{proof}

We now state the main result of this Section, which provides a
performance guarantee for~QCF.
\begin{theorem}[QCF: Probability of $\epsilon$-consensus]
\label{th:QCF} For any $\epsilon>0$, the probability of
$\epsilon$-consensus $T(G,b,\mathbf{\alpha},\epsilon,p,\Delta)$ is
bounded below
\begin{eqnarray}
\label{th:QCF2}
\mathbb{P}\left[\lim_{i\rightarrow\infty}\sup_{1\leq n\leq
N}|\widetilde{x}_{n}(i)-r|<\epsilon\right] & > &
1-\frac{2|\mathcal{M}|\Delta^{2}}{3N^{2}\epsilon^{2}}\sum_{j\geq
0}\alpha^{2}(j)
-\frac{\left[2Nb^{2}+\frac{4|\mathcal{M}|\Delta^{2}}{3N}\sum_{j\geq
0}\alpha^{2}(j)\right]^{1/2}}{p\Delta}\nonumber
\\ & & -\frac{\left(1+N\lambda_{N}(\overline{L})b^{2}\right)\prod_{j\geq
0}(1+g(j))}{1+\frac{p^{2}\Delta^{2}}{2}\lambda_{2}(\overline{L})}
\end{eqnarray}
where $\{g(j)\}_{j\geq 0}$ is defined in eqn.~(\ref{th:QCconv10}).
\end{theorem}
\begin{proof}
It follows from Theorem~\ref{lm:QCF} that
\begin{equation}
\label{th:QCF3}
\lim_{i\rightarrow\infty}\widetilde{x}_{n}(i)=\widetilde{\theta}~\mbox{a.s.},~~\forall
1\leq n\leq N
\end{equation}
The proof then follows from Lemma~\ref{lm:QCF5}.
\end{proof}
 The lower bound on $T(\cdot)$, given by~(\ref{th:QCF2}), is  uniform, in
the sense that it is applicable for all initial states
$\mathbf{x}(0)\in\mathcal{B}$.
Recall the scaled weight sequence $\mathbf{\alpha}_{s}$, given by eqn.~(\ref{weights}). We introduce the zero-rate probability of $\epsilon$-consensus, $T^{z}(G,b,\epsilon,p,\Delta)$  by
\begin{equation}
\label{prop_T2} T^{z}(G,b,\epsilon,p,\Delta) = \lim_{s\rightarrow
0}T(G,b,\mathbf{\alpha}_{s},\epsilon,p,\Delta)
\end{equation}
The next proposition studies the dependence of the $\epsilon$-consensus probability $T(\cdot)$ and of the zero-rate probability $T^{z}((\cdot)$ on the network and algorithm parameters.
\begin{proposition}[QCF: Tradeoffs]
\label{prop_T}
\begin{itemize}[\setlabelwidth{1)}]

\item{\textbf{1)}} \emph{Limiting quantizer.}  For fixed $G,b,\mathbf{\alpha},\epsilon$, we
have
\begin{equation}
\label{prop_T1} \lim_{\Delta\rightarrow
0,~p\Delta\rightarrow\infty}T(G,b,\mathbf{\alpha},\epsilon,p,\Delta)
= 1
\end{equation}
Since, this holds for arbitrary $\epsilon >0$, we note that, as
$\Delta\rightarrow 0,~p\Delta\rightarrow\infty$,
\begin{eqnarray}
\label{prop_T100}
\mathbb{P}\left[\lim_{i\rightarrow\infty}\widetilde{\mathbf{x}}(i)=r\mathbf{1}\right]
& = & \lim_{\epsilon\rightarrow
0}\mathbb{P}\left[\lim_{i\rightarrow\infty}\sup_{1\leq n\leq
N}|\widetilde{x}_{n}(i)-r|<\epsilon\right]\nonumber \\ & = &
\lim_{\epsilon\rightarrow 0}\left[\lim_{\Delta\rightarrow
0,~p\Delta\rightarrow\infty}T(G,b,\mathbf{\alpha},\epsilon,p,\Delta)\right]\nonumber
=1
\end{eqnarray}
In other words, the QCF algorithm leads to a.s. consensus to the
desired average $r$, as $\Delta\rightarrow
0,~p\Delta\rightarrow\infty$. In particular, it shows that the QCF
algorithm is consensus-consistent.

\item{\textbf{2)}} \emph{zero-rate $\epsilon$-consensus probability.}
Then, for fixed $G,b,\epsilon,p,\Delta$, we have
\begin{equation}
\label{prop_T3} T^{z}(G,b,\epsilon,p,\Delta) \geq
1-\frac{\left(2Nb^{2}\right)^{1/2}}{p\Delta}-\frac{1+N\lambda_{N}(\overline{L})b^{2}}{1+\frac{p^{2}\Delta^{2}}{2}\lambda_{2}(\overline{L})}
\end{equation}

\item{\textbf{3)}} \emph{Optimum quantization step-size $\Delta$.} For fixed $G,b,\epsilon,p$, the optimum
quantization step-size $\Delta$, which maximizes the probability
of $\epsilon$-consensus,
$T(G,b,\mathbf{\alpha},\epsilon,p,\Delta)$, is given by
\begin{eqnarray}
\label{prop_T4}
\Delta^{\ast}(G,b,\mathbf{\alpha},\epsilon,p)&=&\arg\inf_{\Delta\geq
0}\left[\frac{2|\mathcal{M}|\Delta^{2}}{3N^{2}\epsilon^{2}}\sum_{j\geq
0}\alpha^{2}(j)+\frac{\left[2Nb^{2}+\frac{4M\Delta^{2}}{3N}\sum_{j\geq
0}\alpha^{2}(j)\right]^{1/2}}{p\Delta}\right.\\
\nonumber
&&\left.\phantom{\frac{\left[2Nb^{2}+\frac{4|\mathcal{M}|\Delta^{2}}{3N}\sum_{j\geq
0}\alpha^{2}(j)\right]^{1/2}}{p\Delta}}
+\frac{\left(1+N\lambda_{N}(\overline{L})b^{2}\right)\prod_{j\geq
0}(1+g(j))}{1+\frac{p^{2}\Delta^{2}}{2}\lambda_{2}(\overline{L})}\right]
\end{eqnarray}
where $\{g(j)\}_{j\geq 0}$ is defined in eqn.~(\ref{th:QCconv10}).
\end{itemize}
\end{proposition}

\begin{proof}
For item~\textbf{2)}, we note that, as $s\rightarrow 0$,
\[
\sum_{j\geq 0}\alpha_{s}^{2}(j)\rightarrow 0,~~\prod_{j\geq
0}(1+g_{s}(j))\rightarrow 1
\]
The rest follows by simple inspection of eqn.~(\ref{th:QCF2}).
\end{proof}
We comment on Proposition~\ref{prop_T}. Item~\textbf{1)} shows
that the algorithm QCF is consensus-consistent, in the sense that
we can achieve arbitrarily good performance by decreasing the
step-size $\Delta$ and the number of quantization levels, $2p+1$,
appropriately. Indeed, decreasing the step-size increases the
precision of the quantized output and increasing $p$ increases the
dynamic range of the quantizer. However, the fact that
$\Delta\rightarrow 0$ but $p\Delta\rightarrow\infty$ implies that
the rate of growth of the number of levels $2p+1$ should be higher
than the rate of decay of $\Delta$, guaranteeing that in the limit
we have asymptotic consensus with probability one.

For interpreting item~\textbf{2)}, we recall the m.s.e.~versus
convergence rate tradeoff for the QC algorithm, studied in
Subsection~\ref{errorQC}. There, we considered a quantizer with a
countably infinite number of output levels (as opposed to the
finite number of output levels in the QCF) and observed that the
m.s.e. can be made arbitrarily small by rescaling the weight
sequence. By Chebyshev's inequality, this would imply, that, for
arbitrary $\epsilon>0$, the probability of $\epsilon$-consensus, i.e., that we get within an $\epsilon$-ball of the desired average,
can be made as close to 1 as we want. However, this occurs at a
cost of the convergence rate, which decreases as the scaling
factor $s$ decreases. Thus, for the QC algorithm, in the limiting
case, as $s\rightarrow 0$, the probability of $\epsilon$-consensus
(for arbitrary $\epsilon>0$) goes to 1; we call ``limiting
probability''  the zero-rate probability of $\epsilon$-consensus,
justifying the m.s.e. vs convergence rate tradeoff.\footnote{Note
that, for both the algorithms, QC and QCF, we can take the scaling
factor, $s$, arbitrarily close to 0, but not zero, so that, these
limiting performance values are not achievable, but we may get
arbitrarily close to them.} Item~\textbf{2)} shows, that, similar
to the QC algorithm, the QCF algorithm exhibits a tradeoff
between probability of $\epsilon$-consensus vs.~the convergence
rate, in the sense that, by scaling (decreasing $s$), the
probability of $\epsilon$-consensus can be increased. However,
contrary to the QC case, scaling will not lead to probability of
$\epsilon$-consensus arbitrarily close to 1, and, in fact, the
zero-rate probability of $\epsilon$-consensus is strictly less
than one, as given by eqn.~(\ref{prop_T3}). In other words, by
scaling, we can make
$T(G,b,\mathbf{\alpha}_{s},\epsilon,p,\Delta)$ as high as
$T^{z}(G,b,\epsilon,p,\Delta)$, but no higher.

We now interpret the lower bound on the zero-rate probability of
$\epsilon$-consensus, $T^{z}(G,b,\epsilon,p,\Delta)$, and show
that the network topology plays an important role in this context.
We note, that, for a fixed number, $N$, of sensor nodes, the only
way the topology enters into the expression of the lower bound
is through the third term on the R.H.S. Then, assuming that,
\[
N\lambda_{N}(\overline{L})b^{2}\gg
1,~~\frac{p^{2}\Delta^{2}}{2}\lambda_{2}(\overline{L})\gg 1
\]
we may use the approximation
\begin{equation}
\label{prop_T6}
\frac{1+N\lambda_{N}(\overline{L})b^{2}}{1+\frac{p^{2}\Delta^{2}}{2}\lambda_{2}(\overline{L})}\approx
\left(\frac{2Nb^2}{p^2\Delta^2}\right)\frac{\lambda_{N}(\overline{L})}{\lambda_{2}(\overline{L})}
\end{equation}
Let us interpret eqn.~(\ref{prop_T6}) in the case, where the
topology is fixed (non-random). Then for all $i$,
$L(i)=\overline{L}=L$. Thus, for a fixed number, $N$, of sensor
nodes, topologies with smaller $\lambda_{N}(L)/\lambda_{2}(L)$,
will lead to higher zero-rate probability of $\epsilon$-consensus
and, hence, are preferable. We note that, in this context, for
fixed $N$, the class of non-bipartite Ramanujan graphs give the
smallest $\lambda_{N}(L)/\lambda_{2}(L)$ ratio, given a constraint
on the number, $M$, of network edges (see~\cite{tsp06-K-A-M}.)

Item~\textbf{3)} shows that, for given graph topology~$G$, initial sensor data, $b$, the link weight sequence~$\mathbf{\alpha}$, tolerance~$\epsilon$, and the number of levels in the quantizer~$p$, the step-size $\Delta$ plays a
significant role in determining the performance. This gives
insight into the design of quantizers to achieve optimal
performance, given a constraint on the number of quantization
levels, or, equivalently, given a bit budget on the communication.

In the next Subsection, we present some numerical studies on the
QCF algorithm, which demonstrate practical implications of the results just discussed.

\vspace{-.5cm}
\subsection{QCF: Numerical Studies}
\label{NumStudQCF} 
\begin{figure}[htb]
\begin{center}
\includegraphics[height=2.0in, width=2.5in]{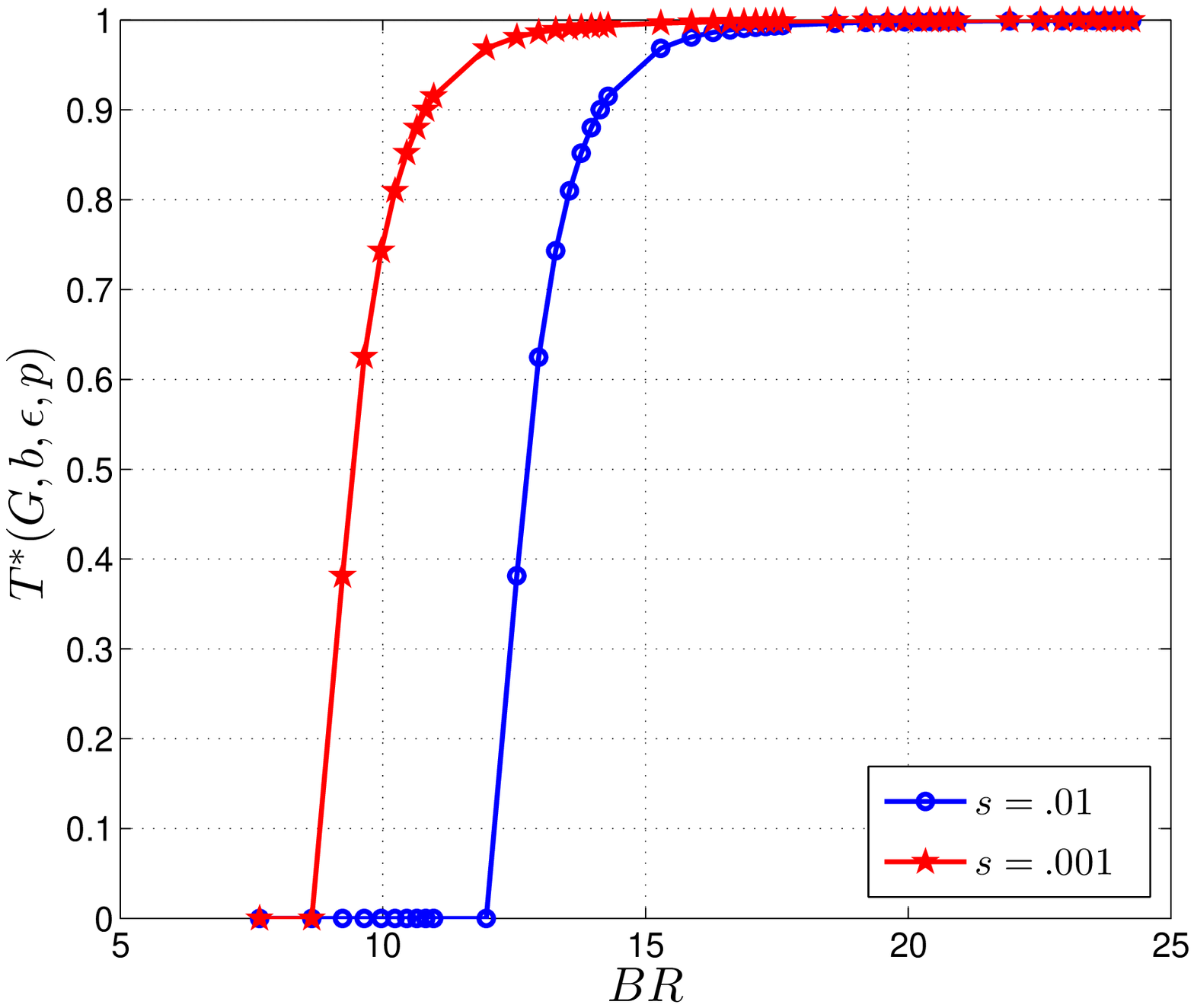}
\includegraphics[height=2.0in, width=2.5in ]{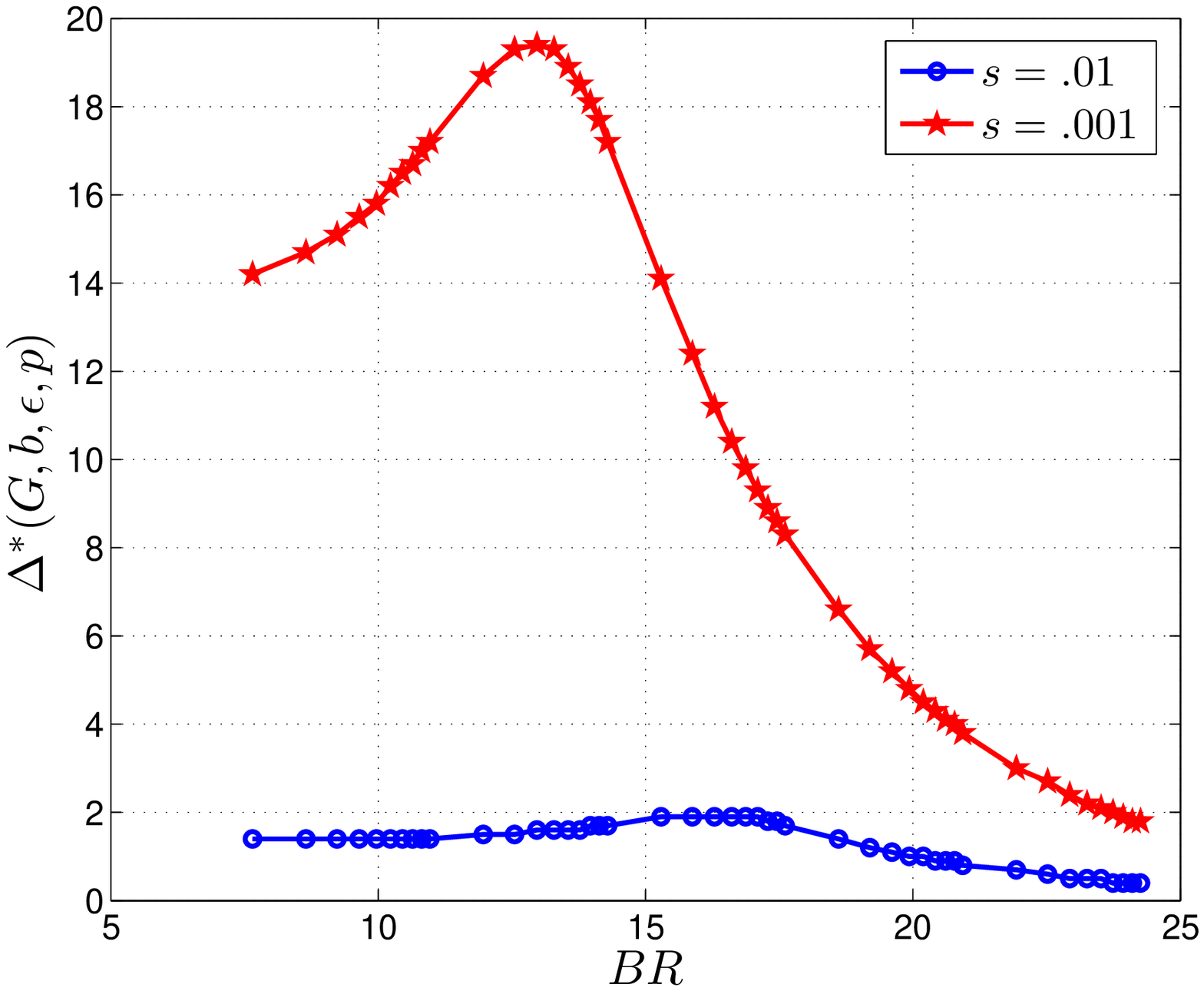}
\caption{Left:
$T^{\ast}(G,b,\mathbf{\alpha}_{s},\epsilon,p)$ vs.~$2p+1$
($BR=\log_{2}(2p+1)$.) Right:
$\Delta^{\ast}(G,b,\mathbf{\alpha}_{s},\epsilon,p)$
vs.~$2p+1$ ($BR=\log_{2}(2p+1)$.) } \label{pr}
\end{center}
\end{figure}

We present a set of numerical studies on the
quantizer step-size optimization problem, considered in
Item~\textbf{3)} of Proposition~\ref{prop_T}. We consider a fixed
(non-random) sensor network of $N=230$ nodes, with communication
topology given by an LPS-II Ramanujan graph
(see~\cite{tsp06-K-A-M}), of degree~6.\footnote{This is a
6-regular graph, i.e., all the nodes have degree~6.} We fix
$\epsilon$ at .05, and take the initial sensor data bound, $b$, to
be 30. We numerically solve the step-size optimization problem
given in~(\ref{prop_T4}) for varying number of levels, $2p+1$.
Specifically, we consider two instances of the optimization
problem: In the first instance, we consider the weight sequence,
$\alpha(i)=.01/(i+1)$, ($s=.01)$, and numerically solve the
optimization problem for varying number of levels. In the second
instance, we repeat the same experiment, with the weight sequence,
$\alpha(i)=.001/(i+1)$, ($s=.001)$. As in eqn.~(\ref{prop_T4}),
$\Delta^{\ast}(G,b,\mathbf{\alpha}_{s},\epsilon,p)$ denotes the
optimal step-size. Also, let
$T^{\ast}(G,b,\mathbf{\alpha}_{s},\epsilon,p)$ be the
corresponding optimum probability of $\epsilon$-consensus.
Fig.~\ref{pr} on the left plots
$T^{\ast}(G,b,\mathbf{\alpha}_{s},\epsilon,p)$ for varying $2p+1$
on the vertical axis, while on the horizontal axis, we plot the
corresponding quantizer bit-rate $BR=\log_{2}(2p+1)$. The two
plots correspond to two different scalings, namely, $s=.01$ and
$s=.001$ respectively. The result is in strict agreement with
Item~\textbf{2)} of Proposition~\ref{prop_T}, and shows that, as
the scaling factor decreases, the probability of
$\epsilon$-consensus increases, till it reaches the zero-rate
probability of $\epsilon$-consensus.

Fig.~\ref{pr} on the right plots
$\Delta^{\ast}(G,b,\mathbf{\alpha}_{s},\epsilon,p)$ for varying
$2p+1$ on the vertical axis, while on the horizontal axis, we plot
the corresponding quantizer bit-rate $BR=\log_{2}(2p+1)$. The two
plots correspond to two different scalings, namely, $s=.01$ and
$s=.001$ respectively. The results are again in strict agreement
to Proposition~\ref{prop_T} and further show that optimizing the
step-size is an important quantizer design problem, because the
optimal step-size value is sensitive to the number of quantization
levels, $2p+1$.

\vspace*{-1cm}
\section{Conclusion}
\label{conclusion} The paper considers distributed average
consensus with quantized information exchange and random
inter-sensor link failures. We add dither to the sensor
states before quantization. We show by stochastic approximation that, when the range of the quantizer is unbounded, the QC-algorithm,
the sensor states achieve a.s.~and m.s.s.~consensus to a random variable whose mean is the desired
average. The variance of this random variable can be
made small by tuning parameters of the algorithm (rate of decay of
the gains), the network topology, and quantizers parameters. When the range of the quantizer is bounded, the QCF-algorithm, a sample path analysis shows that the state vector of the QC-algorithm can be made to remain uniformly bounded with probability arbitrarily close to~$1$. This means that the QCF algorithm achieves $\epsilon$-consensus. We use the bounds that we derive for the probability of large excursions of the sample paths to formulate a quantizer design problem that trades between several quantizer parameters: number of bits (or levels), step size, probability of saturation, and error margin to consensus. A numerical
study illustrates this design problem and several interesting tradeoffs among the design parameters.

\appendices

\renewcommand{\baselinestretch}{1}

%
\section{Proofs of Lemmas~\ref{corr:QC} and~\ref{QC:convmss}}
\label{proofs_res}
\label{imp_res}
Before deriving Lemmas~\ref{corr:QC} and~\ref{QC:convmss},
we present a result from~\cite{kar-moura-ramanan-IT-2008} on a property of real number sequences to be used later, see proof in~\cite{kar-moura-ramanan-IT-2008}.
\begin{lemma}[Lemma 18 in~\cite{kar-moura-ramanan-IT-2008}]
\label{prop} Let the sequences~$\{r_{1}(t)\}_{t\geq 0}$ and
$\{r_{2}(t)\}_{t\geq 0}$ be given by
\begin{equation}
\label{prop1}
r_{1}(t)=\frac{a_{1}}{(t+1)^{\delta_{1}}},~~~~~r_{2}(t)=\frac{a_{2}}{(t+1)^{\delta_{2}}}
\end{equation}
where~$a_{1},a_{2},\delta_{2}\geq 0$ and~$0\leq\delta_{1}\leq 1$.
Then, if~$\delta_{1}=\delta_{2}$ there exists~$K>0$ such that,
for non-negative integers,~$s<t$,
\begin{equation}
\label{prop2}
0\leq\sum_{k=s}^{t-1}\left[\prod_{l=k+1}^{t-1}(1-r_{1}(l))\right]r_{2}(k)\leq
K
\end{equation}
Moreover, the constant~$K$ can be chosen independently of~$s,t$.
 Also, if~$\delta_{1}<\delta_{2}$, then, for arbitrary fixed~$s$,
\begin{equation}
\label{prop3}
\lim_{t\rightarrow\infty}\sum_{k=s}^{t-1}\left[\prod_{l=k+1}^{t-1}(1-r_{1}(l))\right]r_{2}(k)=0
\end{equation}

\end{lemma}

\begin{proof}[Proof of Lemma~\ref{corr:QC}]
Taking expectations (unconditional) on both sides of eqn.~(\ref{th:QCconv9}) we have
\begin{eqnarray}
\label{corr:QC2}
\mathbb{E}\left[V(i+1,\mathbf{x}(i+1))\right]&\leq& \mathbb{E}\left[V(i,\mathbf{x}(i))\right]
-\alpha(i)\mathbb{E}\left[\varphi(i,\mathbf{x}(i))\right]
\nonumber
\\
&+&
\nonumber
g(i)
\left[1+\mathbb{E}\left[V(i,\mathbf{x}(i))\right]\right]
\end{eqnarray}
We also have the following inequalities for all $i$:
\begin{equation}
\label{corr:QC3}
\lambda_{2}(\overline{L})\left\|\mathbf{x}_{\mathcal{C}^{\perp}}\right\|^{2}\leq V(i,\mathbf{x}(i))=\mathbf{x}_{\mathcal{C}^{\perp}}^{T}\overline{L}\mathbf{x}_{\mathcal{C}^{\perp}}\leq\lambda_{N}(\overline{L})\left\|\mathbf{x}_{\mathcal{C}^{\perp}}\right\|^{2}
\end{equation}
\begin{equation}
\label{corr:QC4}
\lambda_{2}^{2}(\overline{L})\left\|\mathbf{x}_{\mathcal{C}^{\perp}}\right\|^{2}\leq \varphi(i,\mathbf{x}(i))=\mathbf{x}_{\mathcal{C}^{\perp}}^{T}\overline{L}^{2}\mathbf{x}_{\mathcal{C}^{\perp}}\leq\lambda_{N}^{2}(\overline{L})\left\|\mathbf{x}_{\mathcal{C}^{\perp}}\right\|^{2}
\end{equation}
From eqns.~(\ref{corr:QC2},\ref{corr:QC3},\ref{corr:QC4}) we have
\begin{equation}
\label{corr:QC5}
\mathbb{E}\left[V(i+1,\mathbf{x}(i+1))\right]\leq \left(1-2\alpha(i)\frac{\lambda_{2}^{2}(\overline{L})}{\lambda_{N}(\overline{L})}+g(i)\right) \mathbb{E}\left[V(i,\mathbf{x}(i))\right] +g(i)
\end{equation}
Choose $0<\varepsilon<\frac{2\lambda_{2}^{2}(\overline{L})}{\lambda_{N}(L)}$ and note that,
the form of $g(i)$ in eqn.~(\ref{th:QCconv10}) and the fact that $\alpha(i)\rightarrow 0$ as $i\rightarrow\infty$ suggests that there exists $i_{\varepsilon}\geq 0$, such that, $\varepsilon\alpha(i)\geq g(i),~~~i\geq i_{\varepsilon}$.
We then have
\begin{equation}
\label{corr:QC7}
\mathbb{E}\left[V(i+1,\mathbf{x}(i+1))\right]\leq \left(1-\left(2\frac{\lambda_{2}^{2}(\overline{L})}{\lambda_{N}(\overline{L})}-\varepsilon\right)\alpha(i)\right) \mathbb{E}\left[V(i,\mathbf{x}(i))\right] +g(i),~~~i\geq i_{\varepsilon}
\end{equation}
Continuing the recursion we have for $i>i_{\varepsilon}$,
\begin{eqnarray}
\label{corr:QC8}
\mathbb{E}\left[V(i,\mathbf{x}(i))\right] & \leq &  \prod_{j=i_{\varepsilon}}^{i-1}
\left(1-\left(2
\frac{\lambda_{2}^{2}(\overline{L})}{\lambda_{N}(\overline{L})}
-\varepsilon\right)\alpha(j)\right)
\mathbb{E}\left[V(i,\mathbf{x}(i_{\varepsilon}))\right]
\\
\nonumber
&+& \sum_{j=i_{\varepsilon}}^{i-1}\left[\left(\prod_{l=j+1}^{i-1}\left(1-\left(2\frac{\lambda_{2}^{2}(\overline{L})}{\lambda_{N}(\overline{L})}-\varepsilon\right)\alpha(l)\right) \right)g(j)\right]
\nonumber
\\
\nonumber
& \leq &
e^{-\left(2
\frac{\lambda_{2}^{2}(\overline{L})}{\lambda_{N}(\overline{L})}
-\varepsilon\right)\sum_{j=i_{\varepsilon}}^{i-1}\alpha(j)}
\mathbb{E}\left[V(i,\mathbf{x}(i_{\varepsilon}))\right]+ \sum_{j=i_{\varepsilon}}^{i-1}\left[\left(
\prod_{l=j+1}^{i-1}\left(1-\left(2
\frac{\lambda_{2}^{2}(\overline{L})}{\lambda_{N}(\overline{L})}
-\varepsilon\right)\alpha(l)\right) \right)g(j)\right]
\end{eqnarray}
where we use $1-a\leq e^{-a}$ for $a\geq 0$. Since the $\alpha(i)$s sum to infinity, we have
\begin{equation}
\label{corr:QC9}
\lim_{i\rightarrow\infty}e^{-\left(2\frac{\lambda_{2}^{2}(\overline{L})}{\lambda_{N}(\overline{L})}-\varepsilon\right)\sum_{j=i_{\varepsilon}}^{i-1}\alpha(j)}=0
\end{equation}
The second term on the R.H.S.~of~(\ref{corr:QC8}) falls under Lemma~\ref{prop} whose second part (eqn.~(\ref{prop3})) implies
\begin{equation}
\label{corr:QC10}
\lim_{i\rightarrow\infty}\sum_{j=i_{\varepsilon}}^{i-1}\left[\left(\prod_{l=j+1}^{i-1}\left(1-\left(2\frac{\lambda_{2}^{2}(\overline{L})}{\lambda_{N}(\overline{L})}-\varepsilon\right)\alpha(l)\right) \right)g(j)\right]=0
\end{equation}
We conclude from eqn.~(\ref{corr:QC8}) that
$\lim_{i\rightarrow\infty}\mathbb{E}\left[V(i,\mathbf{x}(i))\right]=0$.
This with~(\ref{corr:QC3}) implies $\lim_{i\rightarrow \infty}\mathbb{E}\left[\left\|\mathbf{x}_{\mathcal{C}^{\perp}}(i)\right\|^{2}\right]=0$.
From the orthogonality arguments we have for all $i$
\begin{equation}
\label{corr:QC13}
\mathbb{E}\left[\left\|\mathbf{x}(i)-\theta\mathbf{1}\right\|^{2}\right]=\mathbb{E}\left[\left\|\mathbf{x}_{\mathcal{C}}(i)-\theta\mathbf{1}\right\|^{2}\right]+\mathbb{E}\left[\left\|\mathbf{x}_{\mathcal{C}^{\perp}}(i)\right\|^{2}\right]
\end{equation}
The second term in eqn.~(\ref{corr:QC13}) goes to zero by the above, whereas the first term goes to zero by the $\mathcal{L}_{2}$ convergence of the sequence $\{x_{\mbox{\scriptsize{avg}}}(i)\}_{i\geq 0}$ to $\theta$ and the desired m.s.s. convergence follows.
\end{proof}

\begin{proof}[Proof of Lemma~\ref{QC:convmss}]
From~(\ref{corr:QC3},\ref{corr:QC8}), using repeatedly
 $1-a\leq e^{-a}$ for $a\geq0$,
 we have for $i>i_{\varepsilon}$
\begin{eqnarray}
\label{QC:convmss5}
\mathbb{E}\left[\left\|\mathbf{x}_{\mathcal{C}^{\perp}}\right\|^{2}\right]\leq \frac{1}{\lambda_{2}(\overline{L})}
\mathbb{E}\left[V(i,\mathbf{x}(i))\right] & \leq & \frac{1}{\lambda_{2}(\overline{L})}e^{-\left(2\frac{\lambda_{2}^{2}(\overline{L})}{\lambda_{N}(\overline{L})}-\varepsilon\right)\sum_{j=i_{\varepsilon}}^{i-1}\alpha(j)}\mathbb{E}\left[V(i_{\varepsilon},\mathbf{x}(i_{\varepsilon}))\right]+ \nonumber \\ & & \frac{1}{\lambda_{2}(\overline{L})}\sum_{j=i_{\varepsilon}}^{i-1}\left[\left(e^{-\left(2\frac{\lambda_{2}^{2}(\overline{L})}{\lambda_{N}(\overline{L})}-\varepsilon\right)\sum_{l=j+1}^{i-1}\alpha(l)} \right)g(j)\right]
\end{eqnarray}
 From the development in the proof of Theorem~\ref{th:QCavg} we note that
\begin{equation}
\label{QC:convmss6}
\mathbb{E}\left[\left\|\mathbf{x}_{\mathcal{C}}(i)-r\mathbf{1}\right\|^{2}\right]=N^{2}\mathbb{E}\left[\left\|\mathbf{x}_{\mbox{\scriptsize{avg}}}(i)-r\right\|^{2}\right]
\leq\frac{2|\mathcal{M}|\Delta^{2}}{3}\sum_{j=0}^{i-1}\alpha^{2}(j)
\end{equation}
We then arrive at the result by using the equality
\begin{equation}
\label{QC:convmss7}
\left\|\mathbf{x}(i)-r\mathbf{1}\right\|^{2}=\left\|\mathbf{x}_{\mathcal{C}^{\perp}}(i)\right\|^{2}+\left\|\mathbf{x}_{\mathcal{C}}(i)-r\mathbf{1}\right\|^{2},~~~\forall i
\end{equation}
\end{proof}
\vspace*{-1cm}
\section{Proofs of results in Subsection~\ref{samppathbounds}}
\label{app:sampbound}

\begin{proof}[\textbf{Proof of Lemma~\ref{Wsupmart}}]
From eqn.~(\ref{th:QCconv9}) we have
\begin{eqnarray} 
\label{Wsupmart1}
\mathbb{E}\left[V(i+1,\mathbf{x}(i+1))|\mathbf{x}(i)\right]
&\leq&
-\alpha(i)\varphi(i,\mathbf{x}(i))
\nonumber
\\
&+&
\nonumber
g(i)\left[1+V(i,\mathbf{x}(i))\right]+V(i,\mathbf{x}(i))
\end{eqnarray}
We then have
\begin{eqnarray}
\label{Wsupmart2}
\mathbb{E}\left[\left.W(i+1,\mathbf{x}(i+1))\right| \mathcal{F}_{i}\right] &
= &
\mathbb{E}\left[\left.\left(1+V\left(i+1,\mathbf{x}(i+1)\right)\right)\prod_{j\geq
i+1}\left[1+g(j)\right]\right|\mathbf{x}(i)\right]\nonumber \\ & = &
\prod_{j\geq
i+1}\left[1+g(j)\right]\left(\rule{0ex}{2.5ex}1+\mathbb{E}\left[\left.\rule{0ex}{2.2ex}V(i+1,\mathbf{x}(i+1))\right|\mathbf{x}(i)\right]\right)\nonumber
\\ & \leq & \prod_{j\geq
i+1}\left[1+g(j)\right]\left(\rule{0ex}{2.5ex}1-\alpha(i)\varphi(i,\mathbf{x}(i))+g(i)\left[1+V(i,\mathbf{x}(i))\right]+V(i,\mathbf{x}(i))\right)\nonumber
\\ & = & -\alpha(i)\varphi(i,\mathbf{x}(i))\prod_{j\geq
i+1}\left[1+g(j)\right]+\left[1+V(i,\mathbf{x}(i))\right]\prod_{j\geq
i}\left[1+g(j)\right]\nonumber \\ & = &
-\alpha(i)\varphi(i,\mathbf{x}(i))\prod_{j\geq
i+1}\left[1+g(j)\right]+W(i,\mathbf{x}(i))
\end{eqnarray}
Hence $
\mathbb{E}\left[W(i+1,\mathbf{x}(i+1))~|~\mathcal{F}_{i}\right]\leq
W(i,\mathbf{x}(i))$ and the result follows.
\end{proof}

\begin{proof}[\textbf{Proof of Lemma~\ref{lm:sbound1}}]
For any $a>0$ and $i\geq 0$, we have
\begin{equation}
\label{eq:sbound2} \|\mathbf{x}_{\mathcal{C}^{\perp}}(i)\|^{2}>
a~\Longrightarrow~\mathbf{x}(i)^{T}\overline{L}\mathbf{x}(i)\geq
a\lambda_{2}(\overline{L})
\end{equation}
Define the potential function $V\left(i,\mathbf{x}\right)$ as in Theorem~\ref{th:QCconv} and
eqn.~(\ref{th:QCconv4}) and the $W\left(i,\mathbf{x}\right)$ as in~(\ref{Wdef}) in Lemma~\ref{Wsupmart}.
It then follows from eqn.~(\ref{eq:sbound2}) 
 that
\begin{equation}
\label{eq:sbound4} \|\mathbf{x}_{\mathcal{C}^{\perp}}(i)\|^{2}>
a~\Longrightarrow~W(i,\mathbf{x}(i))> 1+a\lambda_{2}(\overline{L})
\end{equation}
By Lemma~\ref{Wsupmart}, the process
$\left(W(i,\mathbf{x}(i)),\mathcal{F}_{i}\right)$ is a
non-negative supermartingale. Then by a maximal inequality for
non-negative supermartingales (see~\cite{Kushner}) we have for
$a>0$ and $i\geq 0$,
\begin{equation}
\label{eq:sbound5} \mathbb{P}\left[\max_{0\leq j\leq
i}W(j,\mathbf{x}(j))\geq
a\right]\leq\frac{\mathbb{E}\left[W(0,\mathbf{x}(0))\right]}{a}
\end{equation}
Also, we note that
\begin{equation}
\label{eq:sbound6} \left\{\sup_{j\geq
0}W(j,\mathbf{x}(j))>a\right\} \Longleftrightarrow \cup_{i\geq
0}\left\{\max_{0\leq j\leq i}W(j,\mathbf{x}(j))> a\right\}
\end{equation}
Since $\left\{\max_{0\leq j\leq i}W(j,\mathbf{x}(j))> a\right\}$
is a non-decreasing sequence of sets in $i$, it follows from the
continuity of probability measures and eqn.~(\ref{eq:sbound4})
\begin{eqnarray}
\label{eq:sbound10}\mathbb{P}\left[\sup_{j\geq
0}\|\mathbf{x}_{\mathcal{C}^{\perp}}(j)\|^{2}>a\right] & = &
\lim_{i\rightarrow\infty}\mathbb{P}\left[\max_{0\leq j\leq
i}\|\mathbf{x}_{\mathcal{C}^{\perp}}(j)\|^{2}>a\right]\\ \nonumber
& \leq & \lim_{i\rightarrow\infty}\mathbb{P}\left[\max_{0\leq
j\leq i}W(j,\mathbf{x}(j))> 1+a\lambda_{2}(\overline{L})\right]\nonumber \\
& \leq &
\lim_{i\rightarrow\infty}\frac{\mathbb{E}\left[W(0,\mathbf{x}(0))\right]}{1+a\lambda_{2}(\overline{L})}\nonumber
=
\frac{\left(1+\mathbf{x}(0)^{T}\overline{L}\mathbf{x}(0)\right)\prod_{j\geq
0}(1+g(j))}{1+a\lambda_{2}(\overline{L})}
\end{eqnarray}
\end{proof}

\begin{proof}[\textbf{Proof of Lemma~\ref{lm:sbound5}}]
It was shown in Theorem~\ref{th:QCavg} that the sequence
$\left\{x_{\mbox{\scriptsize{avg}}}(i)\right\}_{i\geq 0}$ is a martingale. It then follows that the sequence,
$\left\{|x_{\mbox{\scriptsize{avg}}}(i)|\right\}_{i\geq 0}$, is a
non-negative submartingale (see~\cite{Williams}).

The submartingale inequality then states that for $a>0$
\begin{equation}
\label{eq:sbound11} \mathbb{P}\left[\max_{0\leq j\leq
i}|x_{\mbox{\scriptsize{avg}}}(j)|\geq a\right]\leq\frac{
\mathbb{E}\left[|x_{\mbox{\scriptsize{avg}}}(i)|\right]}{a}
\end{equation}
Clearly, from the continuity of probability measures,
\begin{equation}
\label{eq:sbound12} \mathbb{P}\left[\sup_{j\geq
0}|x_{\mbox{\scriptsize{avg}}}(j)|>
a\right]=\lim_{i\rightarrow\infty}\mathbb{P}\left[\max_{0\geq
j\geq i}|x_{\mbox{\scriptsize{avg}}}(j)|> a\right]
\end{equation}
Thus, we have
\begin{equation}
\label{eq:sbound13} \mathbb{P}\left[\sup_{j\geq
0}|x_{\mbox{\scriptsize{avg}}}(j)|>
a\right]\leq\lim_{i\rightarrow\infty}\frac{\mathbb{E}\left[\left|x_{\mbox{\scriptsize{avg}}}(i)\right|\right]}{a}
\end{equation}
(the limit on the right exists because
$x_{\mbox{\scriptsize{avg}}}(i)$ converges in $\mathcal{L}_{1}$.)
 Also, we have from eqn.~(\ref{th:QCavg9}), for all $i$,
\begin{eqnarray}
\label{eq:sbound14}
\mathbb{E}\left[\left|x_{\mbox{\scriptsize{avg}}}(i)\right|\right]
& \leq &
\leq
\left[\mathbb{E}\left[\left|x_{\mbox{\scriptsize{avg}}}(i)\right|^{2}\right]\right]^{1/2}
\leq
\left[x_{\mbox{\scriptsize{avg}}}^{2}(0)+\frac{2|\mathcal{M}|\Delta^{2}}{3N^{2}}\sum_{j\geq
0}\alpha^{2}(j)\right]^{1/2}
\end{eqnarray}
Combining eqns.~(\ref{eq:sbound13},\ref{eq:sbound14}), we have
\begin{equation}
\label{eq:sbound15} \mathbb{P}\left[\sup_{j\geq
0}\left|x_{\mbox{\scriptsize{avg}}}(j)\right|> a\right]\leq
\frac{\left[x_{\mbox{\scriptsize{avg}}}^{2}(0)+\frac{2|\mathcal{M}|\Delta^{2}}{3N^{2}}\sum_{j\geq
0}\alpha^{2}(j)\right]^{1/2}}{a}
\end{equation}
\end{proof}

\begin{proof}[\textbf{Proof of Theorem~\ref{th:sampbound}}]
Since,
$\|\mathbf{x}(j)\|^{2}=Nx_{\mbox{\scriptsize{avg}}}^{2}(i)+\left\|\mathbf{x}_{\mathcal{C}^{\perp}}(j)\right\|^{2}$,
we have
\begin{eqnarray}
\label{th:sampbound2} \mathbb{P}\left[\sup_{j\geq
0}\|\mathbf{x}(j)\|^{2}>a\right] & \leq &
\mathbb{P}\left[\sup_{j\geq
0}N|x_{\mbox{\scriptsize{avg}}}(j)|^{2}>
\frac{a}{2}\right]+\mathbb{P}\left[\sup_{j\geq
0}\|\mathbf{x}_{\mathcal{C}^{\perp}}(j)\|^{2}>\frac{a}{2}\right]\\
\nonumber & = & \mathbb{P}\left[\sup_{j\geq
0}|x_{\mbox{\scriptsize{avg}}}(j)|>
\left(\frac{a}{2N}\right)^{1/2}\right]+\mathbb{P}\left[\sup_{j\geq
0}\|\mathbf{x}_{\mathcal{C}^{\perp}}(i)\|^{2}>\frac{a}{2}\right]
\end{eqnarray}
We thus have from Lemmas~\ref{lm:sbound1} and~\ref{lm:sbound5},
\begin{equation}
\label{th:sampbound3} \mathbb{P}\left[\sup_{j\geq
0}\|\mathbf{x}(j)\|^{2}>a\right]\leq
\frac{\left[x_{\mbox{\scriptsize{avg}}}^{2}(0)+\frac{2|\mathcal{M}|\Delta^{2}}{3N^{2}}\sum_{j\geq
0}\alpha^{2}(j)\right]^{1/2}}{\left(\frac{a}{2N}\right)^{1/2}}+\frac{\left(1+\mathbf{x}(0)^{T}\overline{L}\mathbf{x}(0)\right)\prod_{j\geq
0}(1+g(j))}{1+\left(\frac{a}{2}\right)\lambda_{2}(\overline{L})}
\end{equation}
\end{proof}

\begin{proof}[\textbf{Proof of Corollary~\ref{corr:sampbound}}]
We note that, for $\mathbf{x}(0)\in\mathcal{B}$,
\begin{equation}
\label{corr:sampbound2} x_{\mbox{\scriptsize{avg}}}^{2}(0)\leq
b^{2},~~\mathbf{x}(0)^{T}\overline{L}\mathbf{x}(0)\leq
N\lambda_{N}(\overline{L})b^{2}
\end{equation}
From Theorem~\ref{th:sampbound}, we then get,
\begin{eqnarray}
\label{QCFbound} \mathbb{P}\left[\sup_{1\leq n\leq N,j\geq
0}|x_{n}(j)|>a\right] & \leq & \mathbb{P}\left[\sup_{j\geq
0}\|\mathbf{x}(j)\|>a\right] \nonumber \\
\nonumber
& \leq &
\frac{\left[2Nx_{\mbox{\scriptsize{avg}}}^{2}(0)+\frac{4|\mathcal{M}|\Delta^{2}}{3N}\sum_{j\geq
0}\alpha^{2}(j)\right]^{1/2}}{a}
+\frac{\left(1+\mathbf{x}(0)^{T}\overline{L}\mathbf{x}(0)\right)\prod_{j\geq
0}(1+g(j))}{1+\frac{a^{2}}{2}\lambda_{2}(\overline{L})} \nonumber \\
& \leq &
\frac{\left[2Nb^{2}+\frac{4|\mathcal{M}|\Delta^{2}}{3N}\sum_{j\geq
0}\alpha^{2}(j)\right]^{1/2}}{a}
\nonumber
+\frac{\left(1+N\lambda_{N}(\overline{L})b^{2}\right)\prod_{j\geq
0}(1+g(j))}{1+\frac{a^{2}}{2}\lambda_{2}(\overline{L})}
\end{eqnarray}
\end{proof}


\section{Proofs of Lemma~\ref{lm:QCF5}}
\label{app:3}

\begin{proof}[\textbf{Proof of Lemma~\ref{lm:QCF5}}]
For the proof, consider the sequence $\left\{\mathbf{x}(i)\right\}_{i\geq 0}$
generated by the QC algorithm, with the same initial state
$\mathbf{x}(0)$. Let $\theta$ be the a.s. finite random variable
(see eqn.~\ref{th:QCconv11}) such that
\begin{equation}
\label{lm:QCF2-b}
\mathbb{P}\left[\lim_{i\rightarrow\infty}\mathbf{x}(i)=\theta\mathbf{1}\right]=1
\end{equation}
We note that
\begin{eqnarray}
\label{lm:QCF8}
\mathbb{P}\left[\left|\widetilde{\theta}-r\right|\geq\epsilon\right] & = &
\mathbb{P}\left[\left(\left|\widetilde{\theta}-r\right|\geq\epsilon\right)\cap
\left(\widetilde{\theta}=\theta\right)\right]+
\mathbb{P}\left[\left(\left|\widetilde{\theta}-r\right|\geq\epsilon\right)
\cap\left(\widetilde{\theta}\neq\theta\right)\right] \nonumber \\ & = &
\mathbb{P}\left[\left(\left|\theta-r\right|\geq\epsilon\right)\cap
\left(\widetilde{\theta}=\theta\right)\right]
+\mathbb{P}\left[\left(\left|\widetilde{\theta}-r\right|\geq\epsilon\right)\cap
\left(\widetilde{\theta}\neq\theta\right)\right] \nonumber \\ & \leq &
\mathbb{P}\left[\left|\theta -
r\right|\geq\epsilon\right]+\mathbb{P}\left[\widetilde{\theta}\neq\theta\right]
\end{eqnarray}
From Chebyshev's inequality, we have
\begin{eqnarray}
\label{lm:QCF9} \mathbb{P}\left[\left|\theta -
r\right|\geq\epsilon\right] & \leq &
\frac{\mathbb{E}\left[\left|\theta -
r\right|^{2}\right]}{\epsilon^{2}} \nonumber
\leq
\frac{2|\mathcal{M}|\Delta^{2}}{3N^{2}\epsilon^{2}}\sum_{j\geq
0}\alpha^{2}(j)
\end{eqnarray}
Next, we bound
$\mathbb{P}\left[\widetilde{\theta}\neq\theta\right]$. To this
end, we note that
\begin{eqnarray}
\label{lm:QCF10} \sup_{i\geq 0}\sup_{1\leq n\leq
N}\sup_{l\in\Omega_{n}(i)}\left|x_{l}(i)+\nu_{nl}(i)\right| & \leq
& \sup_{i\geq 0}\sup_{1\leq n\leq
N}\sup_{l\in\Omega_{n}(i)}\left|x_{l}(i)\right|+\sup_{i\geq
0}\sup_{1\leq n\leq
N}\sup_{l\in\Omega_{n}(i)}\left|\nu_{nl}(i)\right| \nonumber
\\ & \leq & \sup_{i\geq 0}\sup_{1\leq n\leq
N}\left|x_{n}(i)\right|+\sup_{i\geq 0}\sup_{1\leq n\leq
N}\sup_{l\in\Omega_{n}(i)}\left|\nu_{nl}(i)\right| \nonumber \\ &
\leq & \sup_{i\geq 0}\sup_{1\leq n\leq
N}\left|x_{n}(i)\right|+\frac{\Delta}{2}
\end{eqnarray}
Then, for any $\delta >0$,
\begin{eqnarray}
\label{lm:QCF11}
\mathbb{P}\left[\widetilde{\theta}\neq\theta\right] & = &
\mathbb{P}\left[\sup_{i\geq 0}\sup_{1\leq n\leq
N}\sup_{l\in\Omega_{n}(i)}\left|x_{l}(i)+\nu_{nl}(i)\right|\geq
\left(p+\frac{1}{2}\right)\Delta\right] \nonumber \\
& \leq &
\mathbb{P}\left[\sup_{i\geq 0}\sup_{1\leq n\leq
N}\left|x_{n}(i)\right|+\frac{\Delta}{2}\geq
\left(p+\frac{1}{2}\right)\Delta\right] \nonumber
\\ & = &
 \mathbb{P}\left[\sup_{i\geq 0}\sup_{1\leq n\leq
N}\left|x_{n}(i)\right|\geq p\Delta\right] \nonumber
\leq
\mathbb{P}\left[\sup_{i\geq 0}\sup_{1\leq n\leq
N}\left|x_{n}(i)\right|> p\Delta-\delta\right] \nonumber \\ & \leq
&
\frac{\left[2Nb^{2}+\frac{4|\mathcal{M}|\Delta^{2}}{3N}\sum_{j\geq
0}\alpha^{2}(j)\right]^{1/2}}{p\Delta-\delta}
+\frac{\left(1+N\lambda_{N}(\overline{L})b^{2}\right)\prod_{j\geq
0}(1+g(j))}{1+\frac{\left(p\Delta-\delta\right)^{2}}{2}\lambda_{2}(\overline{L})}
\end{eqnarray}
where, in the last step, we use eqn.~(\ref{QCFbound}.) Since the
above holds for arbitrary $\delta >0$, we have
\begin{eqnarray}
\label{lm:QCF12}
\mathbb{P}\left[\widetilde{\theta}\neq\theta\right] & \leq &
\lim_{\delta\downarrow
0}\left[\frac{\left[2Nb^{2}+\frac{4|\mathcal{M}|\Delta^{2}}{3N}\sum_{j\geq
0}\alpha^{2}(j)\right]^{1/2}}{p\Delta-\delta}
+\frac{\left(1+N\lambda_{N}(\overline{L})b^{2}\right)\prod_{j\geq
0}(1+g(j))}{1+\frac{\left(p\Delta-\delta\right)^{2}}{2}\lambda_{2}(\overline{L})}\right] \nonumber \\
& = &
\frac{\left[2Nb^{2}+\frac{4|\mathcal{M}|\Delta^{2}}{3N}\sum_{j\geq
0}\alpha^{2}(j)\right]^{1/2}}{p\Delta}
+\frac{\left(1+N\lambda_{N}(\overline{L})b^{2}\right)\prod_{j\geq
0}(1+g(j))}{1+\frac{p^{2}\Delta^{2}}{2}\lambda_{2}(\overline{L})}
\end{eqnarray}
Combining eqns.~(\ref{lm:QCF8},\ref{lm:QCF9},\ref{lm:QCF12}), we
get the result.
\end{proof}

\bibliographystyle{IEEEtran}
\bibliography{IEEEabrv,BibOptEqWeights}

\end{document}